\documentclass[twocolumn,pra,showpacs]{revtex4}

\usepackage{amssymb}
\usepackage{amsfonts}
\usepackage{amsmath}
\usepackage{graphicx}

\begin{document}

\title{Quantum-jumps and photon-statistic in fluorescent systems coupled to
classically fluctuating reservoirs}
\author{Adri\'{a}n A. Budini}
\affiliation{Consejo Nacional de Investigaciones Cient\'{\i}ficas y T\'{e}cnicas, Centro
At\'{o}mico Bariloche, Avenida E. Bustillo Km 9.5, (8400) Bariloche,
Argentina}
\date{\today }

\begin{abstract}
In this paper, we develop a quantum-jump approach for describing the
photon-emission process of single fluorophore systems coupled to complex
classically fluctuating reservoirs. The formalism relies on an open quantum
system approach where the dynamic of the system and the reservoir
fluctuations are described through a density matrix whose evolution is
defined by a Lindblad rate equation. For each realization of the photon
measurement processes it is possible to define a conditional system state
(stochastic density matrix) whose evolution depends on both the photon
detection events and the fluctuations between the configurational states of
the reservoir. In contrast to standard fluorescent systems the
photon-to-photon emission process is not a renewal one, being defined by a
(stochastic) waiting time distribution that in each recording event
parametrically depends on the conditional state. The formalism allows
calculating experimental observables such as the full hierarchy of joint
probabilities associated to the time intervals between consecutive photon
recording events. These results provide a powerful basis for characterizing
different situations arising in single-molecule spectroscopy, such as
spectral fluctuations, lifetime fluctuations, and light assisted processes.
\end{abstract}

\maketitle

\section{Introduction}

A powerful theoretical formalism called the quantum-jump approach \cite%
{breuerbook,plenio,carmichaelbook,zoller,carmichael,dalibard,blatt,hegerplenio,hegerfeldt,heger,beige}
was introduced by the quantum optics community for describing experimental
realizations of single open quantum systems subjected to a continuous
measurement process. Even when only one system is under observation, the
quantum-jump approach allows to define a system state (wave vector or
density matrix), whose dynamic takes into account our change of information
due to the continuous measurement action. Apart from new insights in the
quantum measurement theory, the quantum-jump approach provides an
alternative formalism for characterizing the radiation pattern of single
fluorescent systems driven by a laser field.

While a wide class of quantum optical systems can be studied with the
quantum-jump approach \cite{breuerbook,plenio,carmichaelbook}, it has been
scarcely applied in the context of single-molecule (fluorescence)
spectroscopy (SMS) \cite{barkaiChem,michel,jung}, i.e., in the
characterization of single fluorescent systems coupled to complex host
classically fluctuating environments, such as of those associated to
biological or artificially designed nanoscopic reservoirs. The main task of
SMS is to deduce the underlying environment stochastic dynamic from the
statistical properties of the scattered laser field \cite%
{Orrit,mukamelChem,barkai,Hegerfeldt,wang,brown,xieSchenter,mukamel,Sanda,Xie,osadko}%
. In most of the experiments, the scattered electromagnetic field is
measured with photon detectors. Hence, it can be resolved photon-to-photon.

For direct photon-detection measurement schemes the quantum-jump approach
associate to each photon recording event a sudden disruptive change (wave
vector collapse) in the system state, while in the middle intervals between
consecutive events the--conditional--system evolution is smooth and
non-unitary \cite%
{breuerbook,plenio,carmichaelbook,carmichael,zoller,dalibard}. The formalism
provides a simple technique for calculating and reproducing the photon
recording process. For Markovian dissipative dynamics the emission process
is a \textit{renewal} one, i.e., the statistic of the (random) time
intervals between consecutive photon emissions is always the same, being
defined by a probability distribution called waiting time distribution \cite%
{carmichaelbook,carmichael}.

The main obstacle for applying the quantum-jump approach for modeling SMS
experiments comes from the description of the environment fluctuations. As
in general a full microscopic description is lacking, the complexity of the
environment is taken into account by introducing effective time-dependent
stochastic variables that may modify (parametrize) both the unitary and
dissipative fluorophore evolution. In the context of the quantum-jump
approach, it is not clear how these extra (classical) fluctuations must to
be introduced or consistently interpreted in terms of a continuous
measurement action.

On the basis of stochastic models, the formalism of generalized Bloch
equations \cite{zheng,FLHBrown,zhengbis,he,BarkaiExactMandel} allows to
determining the photon counting probabilities, i.e., the probabilities of
detecting $n$-photons up to a given time. Nevertheless, from that approach
it is not easy to know how the renewal property is broken by the external
fluctuations, neither is known which kind of stochastic dynamic may
reproduce the photon emission process. Then, objects like the hierarchy of
joint probabilities associated to the time intervals between consecutive
photon detections events is also unknown. These statistical objects can be
obtained, for example, from a time average along a single measurement
trajectory [see Eqs.~(\ref{WaitUnoEstacion}) and (\ref{WaitDosEstacion})].

The main goal of this paper is to demonstrate that SMS experiments can be
consistently described in the context of a quantum-jump approach. A general
formalism that allows to characterize the photon-to-photon emission process
for a broad class of environment fluctuations arising in SMS is developed.
In each case, we provide (non-renewal) stochastic processes that reproduce
the statistic of the photon recording events. The average of their
associated dynamic in the system-bath Hilbert space recover the density
matrix evolution. As a central result, we get explicit analytical
expressions for the set of joint probabilities densities defining the
statistics of the time intervals between successive photon recording events.
Therefore, our analysis allows to quantify how and how much the photon
emission process departs from a renewal one.

The formulation of an alternative description of SMS experiments based on a
quantum-jump approach relies on the possibility of describing both the
fluorophore and the environment fluctuations through a density matrix
formalism. In Ref.~\cite{OpenSMS} it was demonstrated that a broad class of
SMS experiments can be studied through an open quantum system approach. The
density matrix evolution is given by a Lindblad rate equation \cite{rate},
which allows to characterize in a unified way both the quantum nature of the
fluorescent system as well as the classical nature of the environment
fluctuations. Based on those results, which are consistent \cite{OpenSMS}
with the formalism of stochastic Bloch equations \cite%
{zheng,FLHBrown,zhengbis,he,BarkaiExactMandel}, we formulate the present
treatment.

We remark that a similar analysis was developed in Ref.~\cite{zumofen}. In
contrast, our present analysis allows getting explicit expressions for the
photon emission statistic, which is also analyzed in the limit of slow and
fast environment fluctuations. Furthermore, an explicit formulation of the
underlying stochastic photon emission process is presented. On the other
hand, our results also clarifies some of the assumptions introduced in
previous author's works \cite{rapid,JPB,luzAssisted} as well as in other
stretched related contributions \cite{petruccione}.

The paper is outlined as follows. In Sec. II, on the basis of the results
developed in Ref.~\cite{OpenSMS}, we define the underlying density matrix
formulation. In Sec. III, we develop the quantum-jump approach. Both, the
stochastic dynamic and the statistical characterization of the photon
emission process are established. In Sec. IV we apply the formalism for the
case in which the measurement apparatus only gives information about the
photon emission events. Different specific cases, such as lifetime
fluctuations and light assisted processes, are analyzed in detail. In
Appendix A we analyze the case of measurements that provide information of
both the photon recording events and about the configurational reservoir
transitions. In Sec. V we provide the conclusions.

\section{Density matrix evolution}

The description of SMS experiments based on a density matrix formalism
relies on the possibility of finding analytically manageable microscopic
interactions able to describe the environment fluctuations as well as their
dynamical influence over the system. In Ref.~\cite{OpenSMS}, following an
argument developed by van Kampen \cite{vanKapenBook}, we modeled the
environment through a set of (effective, coarse grained) macrostates, each
one representing the manifold of quantum bath states that lead to the same
system dynamic. Then, the total microscopic dynamic is written in an
effective Hilbert space defined by the external product of the Hilbert
spaces of the system, the background electromagnetic field, and the
configurational space associated to the bath macrostates. The system is
modeled by a two-level optical transition whose characteristic parameters,
i.e., transition frequency and electric dipole, depend on the state of the
environment. The dielectric constant of its local environment also is
parametrized by the bath macrostates. After tracing out the electromagnetic
field and the configurational states, the density matrix $\rho _{S}(t)$ of
the system can be written as \cite{OpenSMS}%
\begin{equation}
\rho _{S}(t)=\sum\nolimits_{R=1}^{R_{\max }}\rho _{R}(t).  \label{RhoSistema}
\end{equation}%
Each auxiliary state $\rho _{R}(t)$ define the system dynamic \textit{given}
that the reservoir is in the $R$-configurational bath state. $R_{\max }$ is
the number of configurational states. The probability $P_{R}(t)$ that the
environment is in a given state at time $t$ follows from 
\begin{equation}
P_{R}(t)=\mathrm{Tr}_{S}[\rho _{R}(t)],  \label{Prob}
\end{equation}%
where $\mathrm{Tr}_{S}[\cdots ]$ denotes a trace operation in the system
Hilbert space. Therefore, the set of states $\{\rho _{R}(t)\}$ encode both
the system dynamic and the fluctuations of the environment. Their dynamic is
defined by a Lindblad rate equation \cite{rate}%
\begin{eqnarray}
\! \dfrac{d\rho _{R}(t)}{dt}\!\! &=&\!\!\dfrac{-i}{\hbar }[H_{R},\rho
_{R}(t)]\!-\!\gamma _{R}(\{D,\rho _{R}(t)\}_{\!+}\!-\!\mathcal{J}[\rho
_{R}(t)])  \label{LindbladRate} \\
&-& \!\! \sum\limits_{\substack{ R^{\prime }  \\ R^{\prime }\neq R}}\!\!%
\frac{\eta _{R^{\prime }R}}{2}\{A^{\dag }A,\rho
_{R}(t)\}_{+}\!+\!\sum\limits _{\substack{ R^{\prime }  \\ R^{\prime }\neq R 
}}\!\!\eta _{RR^{\prime }}A\rho _{R^{\prime }}(t)A^{\dag }\!.  \notag
\end{eqnarray}
The first line of this equation define the unitary and dissipative system
dynamic given that the bath is in the configurational state $R.$ The
Hamiltonian $H_{R}$ reads%
\begin{equation}
H_{R}=\frac{\hbar \omega _{R}}{2}\sigma _{z}+\frac{\hbar \Omega _{R}}{2}%
(\sigma ^{\dagger }e^{-i\omega _{L}t}+\sigma e^{+i\omega _{L}t}),
\label{Hamiltonian}
\end{equation}%
where%
\begin{equation}
\omega _{R}=(\omega _{0}+\delta \omega _{R}).  \label{SpectralShifts}
\end{equation}%
The upper and lower states of the system are denoted as $\left\vert
+\right\rangle $ and $\left\vert -\right\rangle $ respectively. Its
transition frequency is $\omega _{0}.$ $\sigma _{z}$ is the z-Pauli matrix
in the basis $\{\left\vert +\right\rangle ,\left\vert -\right\rangle \}.$
Then, the contribution $\hbar \omega _{0}\sigma _{z}/2$ defines the bare
system Hamiltonian. The constants $\{\delta \omega _{R}\}$ define the
spectral shifts associated to each bath state. The second contribution in
Eq.~(\ref{Hamiltonian}) introduces the interaction between the system and
the external laser excitation, whose frequency is $\omega _{L}.$ The
operators $\sigma ^{\dag }=\left\vert +\right\rangle \left\langle
-\right\vert $ and $\sigma =\left\vert -\right\rangle \left\langle
+\right\vert $ are the raising and lowering operators acting on system
eigenstates. The Rabi frequencies $\{\Omega _{R}\}$\ measure the strength of
the system-laser coupling for each configurational bath state. The rest of
the system operators appearing in Eq.~(\ref{LindbladRate}) are defined by%
\begin{equation}
D=\sigma ^{\dagger }\sigma /2,\ \ \ \ \ \ \ \ \ \ \ \ \ \mathcal{J}[\bullet
]=\sigma \bullet \sigma ^{\dagger },  \label{DandJ}
\end{equation}%
while $\{\cdots \}_{+}$ denotes an anticonmutation operation. Then, the
contribution proportional to the constant $\gamma _{R}$ defines the natural
decay of the system associated to each reservoir state.

The second line in Eq.~(\ref{LindbladRate}) introduces a coupling (with
rates $\eta _{R^{\prime }R})$ between all the states $\{\rho _{R}(t)\},$
representing the fluctuations (transitions) between the configurational
states of the environment. Depending on the definition of the system
operator $A$ different cases are recovered. When $A=I,$ where $I$ is the
identity operator, the transitions between the configurational states do not
depend on the system state. Hence, the probabilities (\ref{Prob}) are
governed by a classical master equation whose structure follows
straightforwardly from Eq.~(\ref{LindbladRate}). This case allows us to
describe situations such as spectral diffusion processes, conformational
environment fluctuations that affect the natural decay of the system, as
well as single fluorophore systems diffusing in a solution \cite{OpenSMS}.
When $A\neq I,$ the configurational fluctuations are statistically entangled
with the state of the system. Depending on the structure of $A$ different
kind of situations can be described such as for example light assisted
process, where the fluctuations of the bath depend on the external laser
field intensity.

\subsection*{Vectorial representation}

In order to establish a general formulation of the quantum-jump approach, we
introduce a vectorial notation that allow to simplifying the presentation
and calculations. To the configurational bath states we associate a
vectorial space, defined by a basis $\{|R)\}_{R=1}^{R_{\max }},$ with $%
(R|R^{\prime })=\delta _{RR^{\prime }},$ each vector $|R)$ being related to
a different configurational bath state \cite{VectorNotation}. The set of
auxiliary states $\{\rho _{R}(t)\}$ allows us to define the vectors%
\begin{equation}
\left\vert \rho _{t}\right) \equiv \sum_{R}\rho _{R}(t)|R),\ \ \ \ \ \ \ \ \
\ \ \left\vert P_{t}\right) \equiv \sum_{R}\mathrm{Tr}_{S}[\rho _{R}(t)]|R).
\end{equation}%
These two objects encode both the system dynamic and the evolution of the
configurational bath states. In fact,%
\begin{equation}
\rho _{S}(t)=(1|\rho _{t}),\ \ \ \ \ \ \ \ \ \ \ \ P_{R}(t)=(R\left\vert
P_{t}\right) ,  \label{RhoS_P}
\end{equation}%
where we have defined the $R$-vector $(1|\equiv \sum_{R}(R|.$ These
identities follows straightforwardly from Eqs.~(\ref{RhoSistema}) and (\ref%
{Prob}) respectively. The normalization of the system state can be written
as $\mathrm{Tr}_{S}[(1|\rho _{t})]=1,$ while the normalization of the
configurational populations read $(1|P_{t})=1.$

With the vectorial notation, the Lindblad rate equation (\ref{LindbladRate})
can be rewritten as%
\begin{equation}
\frac{d\left\vert \rho _{t}\right) }{dt}=\mathcal{\hat{L}}\left\vert \rho
_{t}\right) .  \label{VectorialLindbladRate}
\end{equation}%
The structure of the matrix of system superoperators $\mathcal{\hat{L}}$
follows from (\ref{LindbladRate}). From now on, with the hat symbol we
denote vectors in the $R-$space whose components are superoperators acting
on the system Hilbert space.

\section{Quantum-jump approach}

Our goal is to characterize the photon emission process associated to the
fluorescent system. Of special interest is to determine how the environment
fluctuations broke the renewal property in successive photon emissions. This
property, for example, can be easily determine from a single experimental
realization by measuring the successive time intervals, $\{\tau
_{i}=t_{i}-t_{i-1}\},$ between consecutive photon recording events
(happening at times $t_{i}$ and $t_{i-1}$). Then, one can define the waiting
time distribution%
\begin{equation}
w_{\infty }^{(1)}(\tau )\equiv \left\langle \delta (\tau -\tau
_{i})\right\rangle _{\mathrm{real}},  \label{WaitUnoEstacion}
\end{equation}%
where $\left\langle \cdots \right\rangle _{\mathrm{real}}$ denotes a time
average along a single realization $[\left\langle f(\tau _{i})\right\rangle
_{\mathrm{real}}=\lim_{t\rightarrow \infty }(1/t)\int_{0}^{t}dt^{\prime
}f(\tau _{i}(t^{\prime }))].$ Consequently, $w_{\infty }^{(1)}(\tau _{1})$
defines the \textit{stationary} probability density of the intervals $\{\tau
_{i}\}.$ It satisfies the normalization $\int_{0}^{\infty }d\tau w_{\infty
}^{(1)}(\tau )=1.$ Similarly, one can define the (stationary) probability
distribution $w_{\infty }^{(2)}(\tau _{2},\tau _{1})$\ for two consecutive
intervals ($\tau _{i}$ and $\tau _{i+1}$), i.e.,%
\begin{equation}
w_{\infty }^{(2)}(\tau _{2},\tau _{1})\equiv \left\langle \delta (\tau
_{2}-\tau _{i+1})\delta (\tau _{1}-\tau _{i})\right\rangle _{\mathrm{real}}.
\label{WaitDosEstacion}
\end{equation}%
It fulfills $\int_{0}^{\infty }d\tau _{2}\int_{0}^{\infty }d\tau
_{1}w_{\infty }^{(2)}(\tau _{2},\tau _{1})=1,$ and the consistency relations 
$\int_{0}^{\infty }d\tau _{2}w_{\infty }^{(2)}(\tau _{2},\tau
_{1})=w_{\infty }^{(1)}(\tau _{1}),$ and $\int_{0}^{\infty }d\tau
_{1}w_{\infty }^{(2)}(\tau _{2},\tau _{1})=w_{\infty }^{(1)}(\tau _{2}).$

By knowing both probability distributions, one can quantify how much the
photon emission process departs from a renewal one. The departure from zero
of the dimensionless parameter%
\begin{equation}
\Lambda (\tau _{2},\tau _{1})\equiv \frac{w_{\infty }^{(2)}(\tau _{2},\tau
_{1})}{w_{\infty }^{(1)}(\tau _{2})w_{\infty }^{(1)}(\tau _{1})}-1,
\label{Lambda}
\end{equation}%
measures the strength of the non-renewal effects induced by the bath
fluctuations. In fact, in absence of fluctuations the system dynamics
becomes Markovian obeying the relation $w_{\infty }^{(2)}(\tau _{2},\tau
_{1})=w_{\infty }^{(1)}(\tau _{2})w_{\infty }^{(1)}(\tau _{1}),$ implying $%
\Lambda (\tau _{2},\tau _{1})=0.$ Nevertheless, we remark that non-Markovian
system dynamic may also lead to renewal emission process \cite%
{zumofen,rapid,JPB}.

The possibility of finding analytical expressions for $w_{\infty
}^{(1)}(\tau _{1})$ and $w_{\infty }^{(2)}(\tau _{2},\tau _{1})$ are one of
the central results of this contribution. We solve this task by extending
the quantum-jump approach on the basis of Eq.~(\ref{VectorialLindbladRate})
[Eq.~(\ref{LindbladRate})].

\subsection{Measurement operators}

The quantum-jump approach relies on a quantum measurement theory \cite%
{breuerbook,plenio,carmichaelbook}. Here, the definition of a measurement
operation must to include both the system and the configurational bath
states. If $|\rho )$ is the state previous to a measurement, the state $%
\mathcal{\hat{M}}_{\mu }|\rho )$ after measurement is%
\begin{equation}
\mathcal{\hat{M}}_{\mu }|\rho )=\frac{\mathcal{\hat{J}}_{\mu }|\rho )}{%
\mathrm{Tr}_{S}[(1|\mathcal{\hat{J}}_{\mu }|\rho )]}.  \label{measurement}
\end{equation}%
The vectorial superoperator $\mathcal{\hat{J}}_{\mu }$ define the
unnormalized transformation of $|\rho )$ due to the measurement action.

When not any measurement is performed over the configurational space, one
must to consider only one superoperator $\mathcal{\hat{M}}_{\mu }$
associated to the photon detection events, $\mu \rightarrow $ \textit{%
photon-detector} [see Eqs.~(\ref{MphotonSelf}) and (\ref{MphotonLigth})].
Nevertheless, we will also consider the existence of extra measurement
channels that may provide information about the configurational states of
the reservoir [see Eqs.~(\ref{MesureSelectiveConfoGENERAL}) and (\ref%
{MUAssisted})]. Hence, Eq.~(\ref{VectorialLindbladRate}) is discomposed as%
\begin{equation}
\frac{d\left\vert \rho _{t}\right) }{dt}=(\mathcal{\hat{D}}+\sum_{\mu }%
\mathcal{\hat{J}}_{\mu })\left\vert \rho _{t}\right) ,  \label{vectorial}
\end{equation}%
where $\mathcal{\hat{D}}\equiv \mathcal{\hat{L}}-\sum_{\mu }\mathcal{\hat{J}}%
_{\mu }.$ In the Markovian case, i.e., when the configurational space is
one-dimensional, the (unique) superoperator $\mathcal{\hat{J}}_{\mu }$ is
related to the wave vector collapse after a photon recording event, while $%
\mathcal{\hat{D}}$ defines the conditional dynamic between consecutive
photon-detections \cite{breuerbook,plenio,carmichaelbook}. Here, the
formalism must also to take into account the fluctuations of the
environment, i.e., the vectorial nature of $\left\vert \rho _{t}\right) $
and the existence of different channels (labeled by $\mu $) that may also
provide information about the transitions between the bath states.

Equation (\ref{vectorial}) provides us the basis for characterizing the
recording process. The following formulation is general, being independent
of both the specific structure of Eq.~(\ref{LindbladRate}) and the
definition of the measurement channels $\{\mathcal{\hat{M}}_{\mu }\}.$
Specific examples are worked out in Section IV and Appendix A.

\subsection{Statistic of the detection events}

The statistics of the successive recording events can be obtained after
writing the system dynamics as an integral over all possible measurement
paths. The evolution Eq.~(\ref{vectorial}) can formally be integrated as%
\begin{equation}
|\rho _{t})=e^{\mathcal{\hat{D}}t}\left\vert \rho _{0}\right) +\sum_{\mu
}\int_{0}^{t}e^{\mathcal{\hat{D}}(t-\tau )}\mathcal{\hat{J}}_{\mu
}\left\vert \rho _{\tau }\right) d\tau ,
\end{equation}%
which can straightforwardly be rewritten in terms of the measurement
operators $\{\mathcal{\hat{M}}_{\mu }\}$ as%
\begin{eqnarray}
|\rho _{t}) &=&P_{0}[t,0;\left\vert \rho _{0}\right) ]\mathcal{\hat{T}}%
(t,0)|\rho _{0})  \label{Solution} \\
&&\!\!\!+\!\sum_{\mu }\!\!\int_{0}^{t}\!\!P_{0}[t,\tau ;\mathcal{\hat{M}}%
_{\mu }|\rho _{\tau })]\mathcal{\hat{T}}(t,\tau )\mathcal{\hat{M}}_{\mu
}|\rho _{\tau })\digamma _{\mu }[|\rho _{\tau })]d\tau .  \notag
\end{eqnarray}%
Here, we have introduced the non-unitary propagator%
\begin{equation}
\mathcal{\hat{T}}(t,\tau )|\rho )\equiv \frac{e^{\mathcal{\hat{D}}(t-\tau
)}|\rho )}{\mathrm{Tr}_{S}[(1|e^{\mathcal{\hat{D}}(t-\tau )}\left\vert \rho
\right) ]},  \label{ConditionalNormalizada}
\end{equation}%
the function 
\begin{equation}
P_{0}[t,\tau ;|\rho )]\equiv \mathrm{Tr}_{S}[(1|e^{\mathcal{\hat{D}}(t-\tau
)}|\rho )],  \label{Survival}
\end{equation}%
and the scalar contribution 
\begin{equation}
\digamma _{\mu }[|\rho )]\equiv \mathrm{Tr}_{S}[(1|\mathcal{\hat{J}}_{\mu
}|\rho )].  \label{sprinkling}
\end{equation}

By associating the propagator $\mathcal{\hat{T}}(t,\tau )$ with the
(vectorial) conditional system dynamic between consecutive recording events
(photon-detections or/and configurational transitions), the first line of
Eq.~(\ref{Solution}) can be interpreted as the contribution of all
measurement realizations where not any detection event happens up to time $%
t. $ Consistently, the weight $P_{0}[t,\tau ;|\rho )]$ must to interpreted
as the corresponding (survival) probability for not having any transition in
the interval $(\tau ,t),$ given that the last one happened at time $\tau ,$
where system state is $|\rho ).$

The second line (integral term) of Eq.~(\ref{Solution}) can be read as the
contribution of all realizations where a measurement event happens at time $%
\tau $ [represented by the action of $\mathcal{\hat{M}}_{\mu }$ on $|\rho
_{\tau })$] and not any detection happen up to time $t,$ which justifies the
presence of $\mathcal{\hat{T}}(t,\tau )$ and the survival probability $%
P_{0}[t,\tau ;\mathcal{\hat{M}}_{\mu }|\rho _{\tau })].$ Consistently, $%
\digamma _{\mu }[|\rho _{\tau })]d\tau $ must to define the probability of
having an event in the $\mu $-detector in the time interval $(\tau ,\tau
+d\tau ).$

By expressing Eq.~(\ref{Solution}) as a sum over all possible measurement
outcomes, the previous statistical interpretation can explicitly be
demonstrated. By writing%
\begin{equation}
|\rho _{t})=\mathcal{\hat{G}}(t)|\rho _{0})=\sum_{n=0}^{\infty }\mathcal{%
\hat{G}}^{(n)}(t)|\rho _{0}),  \label{Unravelling}
\end{equation}%
with $\mathcal{\hat{G}}^{(0)}(t)=P_{0}[t,0,\left\vert \rho _{0}\right) ]%
\mathcal{\hat{T}}(t,0),$ from Eq.~(\ref{Solution}) we get%
\begin{eqnarray}
\mathcal{\hat{G}}^{(n)}(t)\! &=&\!\!\sum_{\mu _{n}\cdots \mu
_{1}}\int_{0}^{t}dt_{n}\cdots \!\int_{0}^{t_{2}}dt_{1}\
P_{n}[t,\{t_{i}\}_{1}^{n},\{\mu _{i}\}_{1}^{n}]  \notag \\
&&\times \mathcal{\hat{T}}(t,t_{n})\mathcal{\hat{M}}_{\mu _{n}}\cdots 
\mathcal{\hat{T}}(t_{2},t_{1})\mathcal{\hat{M}}_{\mu _{1}}\mathcal{\hat{T}}%
(t_{1},0).\ \ \ \ \ \ \   \label{Gene}
\end{eqnarray}%
The weight $P_{n}[t,\{t_{i}\}_{1}^{n},\{\mu _{i}\}_{1}^{n}]$ is defined by 
\begin{widetext}
\begin{equation}
P_{n}[t,\{t_{i}\}_{1}^{n},\{\mu _{i}\}_{1}^{n}]=P_{0}[t,t_{n};\mathcal{\hat{M%
}}_{\mu _{n}}\left\vert \rho _{t_{n}}\right) ]\ w_{\mu _{n}}[t_{n},t_{n-1};%
\mathcal{\hat{M}}_{\mu _{n-1}}|\rho _{t_{n-1}})]\ \cdots \ w_{\mu
_{2}}[t_{2},t_{1};\mathcal{\hat{M}}_{\mu _{1}}|\rho _{t_{1}})]\ w_{\mu
_{1}}[t_{1},0;|\rho _{0})].  \label{Joint}
\end{equation}%
\end{widetext}The intermediate states read%
\begin{equation}
|\rho _{t_{i+1}})=\mathcal{\hat{T}}(t_{i+1},t_{i})\mathcal{\hat{M}}_{\mu
_{i}}|\rho _{t_{i}}),  \label{RhoConditional}
\end{equation}%
with $|\rho _{t_{1}})=\mathcal{\hat{T}}(t_{1},0)|\rho _{0}),$ while%
\begin{equation}
w_{\mu }[t,\tau ;|\rho )]\equiv \mathrm{Tr}_{S}[(1|\mathcal{\hat{J}}_{\mu
}e^{\mathcal{\hat{D}}(t-\tau )}|\rho )].  \label{WaitExplicit}
\end{equation}%
Clearly, $\mathcal{\hat{G}}^{(n)}(t)$ [Eq.~(\ref{Gene})] can be associated
to all trajectories where $n$-detection events happen up to time $t,$ each
one at times $\{t_{i}\}_{i=1}^{i=n}$ in the $\{\mu _{i}\}_{1}^{n}$
detectors. The intermediate evolution between detection events $[\mathcal{%
\hat{M}}_{\mu _{i}}]$ is given by $\mathcal{\hat{T}}(t_{i},t_{i-1}).$
Consistently, $P_{n}[t,\{t_{i}\}_{1}^{n},\{\mu _{i}\}_{1}^{n}]$ [Eq.~(\ref%
{Joint})] defines the probability density of each trajectory. Thus, $w_{\mu
_{i}}[t_{i},t_{i-1};\mathcal{\hat{M}}_{\mu _{i-1}}|\rho _{t_{i-1}})]dt_{i}$
can be read as the probability of having a detection event in the $\mu _{i}$%
-detector in the interval $(t_{i},t_{i}+dt_{i})$ given that the last
detection event happened at time $t_{i-1}$ in the $\mu _{i-1}$-detector, not
happening any event inside the interval $(t_{i-1},t_{i}).$

By using the normalization of the vectorial state, $(d/dt)\mathrm{Tr}%
_{S}[(1\left\vert \rho _{t}\right) ]=0,$ from Eq.~(\ref{vectorial}) it
follows the relation $\mathrm{Tr}_{S}[(1|\mathcal{\hat{D}}|\bullet
)]=-\sum_{\mu }\mathrm{Tr}_{S}[(1|\mathcal{\hat{J}}_{\mu }|\bullet )].$
Then, Eq.~(\ref{Survival}) can alternatively be written as%
\begin{equation}
P_{0}[t,\tau ;\left\vert \rho \right) ]=1-\sum_{\mu }\int_{0}^{t}w_{\mu
}[t,\tau ;|\rho )]d\tau .
\end{equation}%
With this relation, we notice that Eq.~(\ref{Joint}) has the same structure
than a renewal process, i.e., there exist a probability distribution
(waiting time distribution, $w_{\mu }[t,\tau ;|\rho )]$) that define the
statistic of the time interval between consecutive detection events.
Nevertheless, here the waiting time distribution has a functional dependence
on the system state posterior to a detection event $[\mathcal{\hat{M}}_{\mu
}|\rho )],$ which broke the renewal property.

By writing the states $|\rho _{t_{i}+1})$ [Eq.~(\ref{RhoConditional})] as%
\begin{equation}
|\rho _{t_{i+1}})=\frac{e^{\mathcal{\hat{D}(}t_{i+1}-t_{i})}\mathcal{\hat{J}}%
_{\mu _{i}}|\rho _{t_{i}})}{\mathrm{Tr}_{S}[(1|e^{\mathcal{\hat{D}(}%
t_{i+1}-t_{i})}\mathcal{\hat{J}}_{\mu _{i}}|\rho _{t_{i}})]},
\end{equation}%
the $n$-joint probability density (\ref{Joint}) can be rewritten as%
\begin{eqnarray}
P_{n}[t,\{t_{i}\}_{1}^{n},\{\mu _{i}\}_{1}^{n}]\! &=&\!\mathrm{Tr}_{S}[(1|e^{%
\mathcal{\hat{D}}(t-t_{n})}\mathcal{\hat{J}}_{\mu _{n}}\cdots \mathcal{\hat{J%
}}_{\mu _{2}}e^{\mathcal{\hat{D}}(t_{2}-t_{1})}  \notag \\
&&\times \mathcal{\hat{J}}_{\mu _{1}}e^{\mathcal{\hat{D}}t_{1}}\left\vert
\rho _{0}\right) ].  \label{JointCorta}
\end{eqnarray}%
This expression recovers the result of Ref.~\cite{zumofen}. Notice that its
structure is similar to that obtained in the context of a photon measurement
theory \cite{breuerbook,plenio,carmichaelbook}. Nevertheless, here the
underlying trajectories are vectorial and depend on the extra parameters $%
\mu _{i},$ $i=1\cdots n.$

The probabilities $P_{n}(t)$ of having $n$-detection events up to time $t$
can be obtained by integrating the joint probabilities densities $%
P_{n}[t,\{t_{i}\}_{1}^{n},\{\mu _{i}\}_{1}^{n}]$ over all possible detection
paths%
\begin{equation}
P_{n}(t)=\sum_{\mu _{n}\cdots \mu _{1}}\int_{0}^{t}dt_{n}\cdots
\int_{0}^{t_{2}}dt_{1}\ P_{n}[t,\{t_{i}\}_{1}^{n},\{\mu _{i}\}_{1}^{n}].
\end{equation}%
In the context of SMS, objects of this kind are usually characterized
through a generating function approach based on a stochastic Bloch equation 
\cite{zheng,FLHBrown,zhengbis,he,BarkaiExactMandel}. Then, while previous
approaches are able to get these objects, the present treatment also allows
us to get the underlying joint statistic defined by Eq.~(\ref{Joint}).

\subsection{Stationary waiting time distributions}

The joint probability density Eq.~(\ref{Joint}) is one of the central
results of this section. It completely characterizes the statistic of the
recording events. It can experimentally be determine from an ensemble
average over measurement realizations having $n$-detection events in the
interval $(0,t).$ Nevertheless, the stationary waiting time distributions
Eqs.~(\ref{WaitUnoEstacion}) and (\ref{WaitDosEstacion}) are defined by a
time average along a single measurement realization. For ergodic environment
fluctuations, objects of this nature can be studied by describing the
measurement process after happening an infinite number of recording events
and that an infinite time elapsed since the initial condition, $\left\vert
\rho _{0}\right) .$ In Appendix B, we show that in that limit Eq.~(\ref%
{Joint}) remains valid under the replacement%
\begin{equation}
\left\vert \rho _{0}\right) \rightarrow \mathcal{\hat{M}}|\rho _{\infty }),
\end{equation}%
where $|\rho _{\infty })$ corresponds to the stationary state%
\begin{equation}
|\rho _{\infty })\equiv \lim_{t\rightarrow \infty }|\rho _{t}).
\label{RhoInfinity}
\end{equation}%
It comes forth because a time averaging procedure can only provides
information about stationary observables. The measurement operator $\mathcal{%
\hat{M}}$ is defined by%
\begin{equation}
\mathcal{\hat{M}}|\rho )\equiv \frac{\mathcal{\hat{J}}|\rho )}{\mathrm{Tr}%
_{S}[(1|\mathcal{\hat{J}}|\rho )]},\ \ \ \ \ \ \ \ \ \ \mathcal{\hat{J}}%
\equiv \sum_{\mu }\mathcal{\hat{J}}_{\mu },  \label{EME}
\end{equation}%
and takes into account the happening of an arbitrary measurement event in
the long time regime. With these definitions, from Eq.~(\ref{Joint}) we
introduce the first stationary waiting time distribution%
\begin{eqnarray}
w_{\infty }^{(1)}(\tau ,\mu ) &\equiv &w_{\mu }[\tau ,0;\mathcal{\hat{M}}%
|\rho _{\infty })]  \label{waitST1} \\
&=&\mathrm{Tr}_{S}[(1|\mathcal{\hat{J}}_{\mu }e^{\mathcal{\hat{D}}\tau }%
\mathcal{\hat{M}}|\rho _{\infty })],  \notag
\end{eqnarray}%
as well as the second order stationary waiting time distribution%
\begin{eqnarray}
w_{\infty }^{(2)}\!(\tau _{2},\mu _{2};\tau _{1},\mu _{1})\!\! &\equiv
&\!\!w_{\mu _{2}}[\tau _{2}+\tau _{1},\tau _{1};\mathcal{\hat{M}}_{\mu
_{1}}\!|\rho _{\tau _{1}})]  \label{waitST2} \\
&&\times w_{\mu _{1}}[\tau _{1},0;\mathcal{\hat{M}}\!|\rho _{\infty })] 
\notag \\
\!\! &=&\!\!\mathrm{Tr}_{S}[(1|\mathcal{\hat{J}}_{\mu _{2}}\!e^{\mathcal{%
\hat{D}}\tau _{2}}\!\mathcal{\hat{J}}_{\mu _{1}}\!e^{\mathcal{\hat{D}}\tau
_{1}}\mathcal{\hat{M}}\!|\rho _{\infty })].  \notag
\end{eqnarray}%
Higher objects, $w_{\infty }^{(n)}\![\{\tau _{i}\}_{1}^{n},\!\{\mu
_{i}\}_{1}^{n}],$ can be written in a similar way. They define, in the
stationary regime, the probability density of the time intervals $\{\tau
_{i}\}_{1}^{n}$ between successive recording events happening in the $\{\mu
_{i}\}_{1}^{n}$ detectors. When the measurement process only involves a
photon detector apparatus, $\mu \rightarrow $photon-detector, Eqs.(\ref%
{waitST1}) and (\ref{waitST2}) allow to get analytical expressions for the
distributions (\ref{WaitUnoEstacion}) and (\ref{WaitDosEstacion})
respectively (see Section IV).

\subsection{Stochastic density matrix evolution}

From the previous analysis, we obtained the recording event statistics
associated to the density matrix evolution Eq.~(\ref{vectorial}). The
quantum-jump approach also allows building up the underlying stochastic
dynamics that reproduce that statistic. The key ingredient is the definition
of a stochastic process developing in the system Hilbert space and whose
realizations can be mapped with the realizations of the measurement
apparatus signals. The average over realizations must to recover the system
density matrix evolution. Then, in the present context we search for the
definition of a stochastic vector $|\rho _{t}^{\mathrm{st}}),$ such that $%
\overline{|\rho _{t}^{\mathrm{st}})}=|\rho _{t}),$ where $|\rho _{t})$ is
defined by the evolution (\ref{vectorial}). From now on, the overbar denotes
(ensemble) averaging over realizations.

Based on the path integral solution obtained previously [Eq.~(\ref%
{Unravelling})], the stochastic evolution can be written as a piecewise
deterministic processes \cite{breuerbook}%
\begin{equation}
\frac{d}{dt}|\rho _{t}^{\mathrm{st}})=[\mathcal{\hat{D}}-\mathrm{Tr}_{S}(1|%
\mathcal{\hat{D}}|\rho _{t}^{\mathrm{st}})]|\rho _{t}^{\mathrm{st}%
})+\sum_{\mu }(\mathcal{\hat{M}}_{\mu }-1)|\rho _{t}^{\mathrm{st}})\frac{%
dN_{t}^{\mu }}{dt}.  \label{stochastic}
\end{equation}%
Here, the deterministic non-linear term [first contribution on the r.h.s.]
corresponds to the \textit{conditional} evolution in the intervals between
consecutive measurements events, i.e., the dynamics defined by Eq.~(\ref%
{ConditionalNormalizada}). On the other hand, the second term introduces the
disruptive changes in the vectorial state after a measurement event, i.e., $%
|\rho _{t}^{\mathrm{st}})\rightarrow \mathcal{\hat{M}}_{\mu }|\rho _{t}^{%
\mathrm{st}}).$ Consistently, the noisy terms are defined by $dN_{t}^{\mu
}/dt\equiv \sum\nolimits_{k}\delta (t-t_{k}^{\mu }),$ where $t_{k}^{\mu }$
are the times where a measurement event happens in the $\mu $-detector. By
denoting with $N_{t}^{\mu }$ the number of detections events up to time $t,$
it follows the alternative definition $dN_{t}^{\mu }=(N_{t+dt}^{\mu
}-N_{t}^{\mu }),$ i.e., $dN_{t}^{\mu }$ are the increments of the (Poisson)
process $N_{t}^{\mu }$ \cite{breuerbook}. In agreement with the previous
analysis, their average must to recover Eq.~(\ref{sprinkling}), i.e.,%
\begin{equation}
\overline{dN_{t}^{\mu }}=\mathrm{Tr}_{S}[(1|\mathcal{\hat{J}}_{\mu }|\rho
_{t})]dt=\digamma _{\mu }[|\rho _{t})]dt.
\end{equation}

By using the property $dN_{t}^{\mu }dN_{t}^{\mu ^{\prime }}=\delta _{\mu \mu
^{\prime }}dN_{t}^{\mu },$ which implies that a simultaneous detection in
two different measurement apparatus is never observed and that $(dN_{t}^{\mu
})^{k}=dN_{t}^{\mu },$ in Appendix C we show that Eq.~(\ref{vectorial}) is
recovered after averaging Eq.~(\ref{stochastic}) over realizations.

The realizations associated to Eq. (\ref{stochastic}) can be easily
determine after providing a recipe for calculating the random times where
the detection events happen. Their numerical calculation relies on
evaluating the statistical objects introduced in Eq.~(\ref{Solution}) along
each trajectory. Given that the system is in the state $|\rho _{t}^{\mathrm{%
st}}),$ the quantity $\digamma _{\mu }[|\rho _{t}^{\mathrm{st}})]dt$ [Eq.~(%
\ref{sprinkling})] gives the probability of having an event in the $\mu $%
-detector in the time interval $(t,t+dt).$ This quantity defines an
infinitesimal time step algorithm (see Appendix D). In a similar way, $%
P_{0}[t,t^{\prime };\mathcal{\hat{M}}_{\mu }|\rho _{t^{\prime }}^{\mathrm{st}%
})]$ [Eq.~(\ref{Survival})] define the survival probability for the next
detection event (at time $t$) given that a $\mu $-detection event happened
at time $t^{\prime }.$ This object allows to defining a finite time step
algorithm (see Appendix D).

Independently of the method (algorithm) used to determine the times of the
recording events (transitions), given that at time $t$ a measurement
happens, $|\rho _{t}^{\mathrm{st}})\rightarrow \mathcal{\hat{M}}_{\mu }|\rho
_{t}^{\mathrm{st}}),$ each transformation $\mathcal{\hat{M}}_{\mu }$ [Eq.~(%
\ref{measurement})] must be chosen with probability%
\begin{equation}
\mathrm{t}_{\mu }(t)\equiv \frac{\digamma _{\mu }[|\rho _{t}^{\mathrm{st}})]%
}{\sum_{\mu ^{\prime }}\digamma _{\mu ^{\prime }}[|\rho _{t}^{\mathrm{st}})]}%
=\frac{\mathrm{Tr}_{S}[(1|\mathcal{\hat{J}}_{\mu }|\rho _{t}^{\mathrm{st}})]%
}{\sum_{\mu ^{\prime }}\mathrm{Tr}_{S}[(1|\mathcal{\hat{J}}_{\mu ^{\prime
}}|\rho _{t}^{\mathrm{st}})]},  \label{pu}
\end{equation}%
which satisfy $\sum_{\mu }\mathrm{t}_{\mu }(t)=1.$ This rule corresponds to
a selective measurement of the set of $\mu $-observables \cite{breuerbook}.
Between successive recording events, the evolution of $|\rho _{t}^{\mathrm{st%
}})$ is deterministic and defined by Eq.~(\ref{ConditionalNormalizada}).

Through the relations%
\begin{equation}
\rho _{S}^{\mathrm{st}}(t)\equiv (1|\rho _{t}^{\mathrm{st}}),\ \ \ \ \ \ \ \
\ \ |P_{t}^{\mathrm{st}})\equiv \mathrm{Tr}_{S}[|\rho _{t}^{\mathrm{st}})],
\label{SystemConfiSTOCH}
\end{equation}%
the vectorial state $|\rho _{t}^{\mathrm{st}})$ provide a stochastic
representation of both the system density matrix [Eq.~(\ref{RhoSistema})], $%
\overline{\rho _{S}^{\mathrm{st}}(t)}=\rho _{S}(t),$ and the occupation of
the configurational bath states [Eq.~(\ref{Prob})], $(R\overline{\left\vert
P_{t}^{\mathrm{st}}\right) }=P_{R}(t).$ In contrast with the standard
quantum-jump approach, in general it is not possible to get a simple
dynamical evolution for $\rho _{S}^{\mathrm{st}}(t)$ [or to $(R|P_{t}^{%
\mathrm{st}})$]. In fact, here the formalism relies on the vectorial nature
of $|\rho _{t}^{\mathrm{st}})$ [however see also Appendix A].

\subsection{Non-renewal recording realizations}

The trajectories associated to Eq.~(\ref{stochastic}) allow us to
establishing a simple scheme for understanding the \textit{non-renewal}
nature of the recording process. In fact, its underlying structure is
similar to that of a renewal one. Given that the last event happened at time 
$t^{\prime }$ in the $\mu $-detector, the random time $t$ for the next event
is defined by a waiting time distribution $w_{\mathrm{st}}(t,t^{\prime },\mu
),$ which read 
\begin{subequations}
\label{WestocasticaPoDef}
\begin{eqnarray}
w_{\mathrm{st}}(t,t^{\prime },\mu ) &\equiv &-\frac{d}{dt}P_{0}[t,t^{\prime
};\mathcal{\hat{M}}_{\mu }|\rho _{t^{\prime }}^{\mathrm{st}})], \\
&=&-\mathrm{Tr}_{S}[(1|\mathcal{\hat{D}}e^{\mathcal{\hat{D}}(t-t^{\prime })}%
\mathcal{\hat{M}}_{\mu }|\rho _{t^{\prime }}^{\mathrm{st}})].
\end{eqnarray}%
By using the relation $\mathrm{Tr}_{S}[(1|\mathcal{\hat{D}}|\bullet
)]=-\sum_{\mu }\mathrm{Tr}_{S}[(1|\mathcal{\hat{J}}_{\mu }|\bullet )],$ it
follows 
\end{subequations}
\begin{equation}
w_{\mathrm{st}}(t,t^{\prime },\mu )=\mathrm{Tr}_{S}[(1|\mathcal{\hat{J}}e^{%
\mathcal{\hat{D}}(t-t^{\prime })}\mathcal{\hat{M}}_{\mu }|\rho _{t^{\prime
}}^{\mathrm{st}})],  \label{Wst_mu}
\end{equation}%
where $\mathcal{\hat{J}}=\sum_{\mu }\mathcal{\hat{J}}_{\mu }$ [Eq.~(\ref{EME}%
)]. At time $t,$ $|\rho _{t^{\prime }}^{\mathrm{st}})$ is updated with the
conditional evolution Eq.~(\ref{ConditionalNormalizada}) and the new
recording event is selected with the probabilities (\ref{pu}). The next
events follow from the same rule (see Appendix D). The average over
realizations recover the statistics defined by Eq.~(\ref{Joint}).

The departure of the recording realizations with respect to a renewal
process comes from the dependence of $w_{\mathrm{st}}(t,t^{\prime },\mu )$
on $|\rho _{t^{\prime }}^{\mathrm{st}}).$ Only if $\mathcal{\hat{M}}_{\mu
}|\rho _{t^{\prime }}^{\mathrm{st}})$ is independent of $|\rho _{t^{\prime
}}^{\mathrm{st}})$ one get a renewal recording process. Nevertheless, in
general this does not happen, implying that $w_{\mathrm{st}}(t,t^{\prime
},\mu )$ randomly change between successive events. Then, in contrast with a
renewal process, here the successive events are defined by a stochastic
waiting time distribution that parametrically depends on the vectorial state 
$|\rho _{t^{\prime }}^{\mathrm{st}}).$ Finally, we notice that $w_{\mathrm{st%
}}(t,t^{\prime },\mu )$ can consistently be written as $w_{\mathrm{st}%
}(t,t^{\prime },\mu )=\sum_{\tilde{\mu}}w_{\tilde{\mu}}[t,t^{\prime };%
\mathcal{\hat{M}}_{\mu }|\rho _{t^{\prime }}^{\mathrm{st}})],$ where $w_{\mu
}[t,t^{\prime };\rho )]$ is defined by Eq.~(\ref{WaitExplicit}).

\section{Photon emission measurements}

In the previous sections, we developed a general theory that allows to
characterizing the measurement processes associated to a broad class of
physical situation arising in SMS. The theory depends on which kind of
measurement process is performed over both the system and the
configurational states. In this section, we analyze the situation where
there exists only one measurement process defined by a photon detector
apparatus coupled to the scattered electromagnetic field. This is the
standard situation in SMS, where any direct information about the
configurational space is unavailable. Then, the parameter $\mu $ only
includes one term corresponding to the photon detector. Furthermore, our
formalism is able to describe different kind of environmental fluctuations.
First, we analyze the case of self-fluctuating environments, i.e., when the
transitions between the configurational states do not depend on the state of
the system. As a second leading case, we analyze environmental fluctuations
that depend on the intensity of the laser excitation.

\subsection{Self-fluctuating environments}

This case is covered by Eq.~(\ref{LindbladRate}) by taking $A=I,$%
\begin{eqnarray}
\dfrac{d\rho _{R}(t)}{dt} &=&\dfrac{-i}{\hbar }[H_{R},\rho _{R}(t)]-\gamma
_{R}(\{D,\rho _{R}(t)\}_{+}-\mathcal{J}[\rho _{R}(t)])  \notag \\
&&-\sum\limits_{R^{\prime }}\phi _{R^{\prime }R}\rho
_{R}(t)+\sum\limits_{R^{\prime }}\phi _{RR^{\prime }}\rho _{R^{\prime }}(t).
\label{LindbladSelf}
\end{eqnarray}%
For notational consistency we take $\eta _{RR^{\prime }}\rightarrow \phi
_{RR^{\prime }}$ \cite{OpenSMS}. From Eq.~(\ref{LindbladSelf}), the
evolution of populations Eq.~(\ref{Prob}) is given by%
\begin{equation}
\frac{d}{dt}P_{R}(t)=-\sum\limits_{R^{\prime }}\phi _{R^{\prime
}R}P_{R}(t)+\sum\limits_{R^{\prime }}\phi _{RR^{\prime }}P_{R^{\prime }}(t).
\label{ClassicalPconfo}
\end{equation}%
Hence, the stochastic dynamics between the configurational states is
governed by a classical master equation that does not depend on the state of
the system. This case allow to describe processes such as spectral
fluctuations, life time fluctuations, and molecules diffusing in a solution.

\subsubsection{Photon measurement operator}

The measurement operator, Eq.~(\ref{measurement}), must to take into account
all contributions that, independently of the $R$-state of the reservoir,
lead to a photon emission. Then, from Eq.~(\ref{LindbladSelf}), we write ($%
\mu \rightarrow \mathrm{ph}$) $[|\rho )=\sum_{R}|R)\rho _{R}]$%
\begin{equation}
\mathcal{\hat{M}}_{\mathrm{ph}}|\rho )=\frac{\mathcal{\hat{J}}_{\mathrm{ph}%
}|\rho )}{\mathrm{Tr}_{S}[(1|\mathcal{\hat{J}}_{\mathrm{ph}}|\rho )]}=\frac{%
\sum_{R}\gamma _{R}|R)\ \sigma \rho _{R}\sigma ^{\dagger }}{\sum_{R^{\prime
}}\gamma _{R^{\prime }}\mathrm{Tr}_{S}[\sigma ^{\dagger }\sigma \rho
_{R^{\prime }}]}.  \label{MphotonSelf}
\end{equation}%
Notice that each contribution in the sum corresponds to the standard
definition arising in Markovian fluorescent systems \cite%
{breuerbook,plenio,carmichaelbook}, i.e., $\mathcal{M}_{\mathrm{ph}}\rho
=\sigma \rho \sigma ^{\dagger }/\mathrm{Tr}_{S}[\sigma ^{\dagger }\sigma
\rho ].$ The vectorial superoperator $\mathcal{\hat{D}}$ [Eq.~(\ref%
{vectorial})] here is defined from%
\begin{equation}
\mathcal{\hat{D}}=\mathcal{\hat{L}}-\mathcal{\hat{J}}_{\mathrm{ph}},
\label{DConfo}
\end{equation}%
where $\mathcal{\hat{L}}$ follows from Eqs.~(\ref{VectorialLindbladRate})
and (\ref{LindbladSelf}), while $\mathcal{\hat{J}}_{\mathrm{ph}}$ from Eq.~(%
\ref{MphotonSelf}). Alternatively, $\mathcal{\hat{D}}$ can be explicitly
defined through the non-unitary evolution generated by it%
\begin{equation}
\frac{d}{dt}(R|\rho _{t}^{\mathrm{u}})=(R|\mathcal{\hat{D}}|\rho _{t}^{%
\mathrm{u}}),  \label{UnNormalized}
\end{equation}%
where the index $\mathrm{u}$ say us that the auxiliary vector $|\rho _{t}^{%
\mathrm{u}})$ is not normalized to one. In fact, its norm is related to the
survival probability Eq.~(\ref{Survival}). By denoting $\rho _{R}^{\mathrm{u}%
}(t)=(R|\rho _{t}^{\mathrm{u}}),$ we get%
\begin{eqnarray}
\dfrac{d\rho _{R}^{\mathrm{u}}(t)}{dt} &=&\dfrac{-i}{\hbar }[H_{R},\rho
_{R}^{\mathrm{u}}(t)]-\gamma _{R}\{D,\rho _{R}^{\mathrm{u}}(t)\}_{+}  \notag
\\
&&-\sum\limits_{R^{\prime }}\phi _{R^{\prime }R}\rho _{R}^{\mathrm{u}%
}(t)+\sum\limits_{R^{\prime }}\phi _{RR^{\prime }}\rho _{R^{\prime }}^{%
\mathrm{u}}(t).  \label{RhoCondConfo}
\end{eqnarray}%
Having the definition of the superoperators $\mathcal{\hat{M}}_{\mathrm{ph}}$
and $\mathcal{\hat{D}}$ we can apply the theory developed in previous
section.

\subsubsection{Stochastic dynamics}

The dynamic of the stochastic state $|\rho _{t}^{\mathrm{st}})$ follows from
Eq.~(\ref{stochastic}). The disruptive transformation associated to a photon
detection event, $|\rho _{t}^{\mathrm{st}})\rightarrow \mathcal{\hat{M}}_{%
\mathrm{ph}}|\rho _{t}^{\mathrm{st}}),$ from the expression (\ref%
{MphotonSelf}), and by using that $\sigma =\left\vert -\right\rangle
\left\langle +\right\vert ,$ $\sigma ^{\dagger }=\left\vert +\right\rangle
\left\langle -\right\vert ,$ can explicitly be written as%
\begin{equation}
|\rho _{t}^{\mathrm{st}})\rightarrow \mathcal{\hat{M}}_{\mathrm{ph}}|\rho
_{t}^{\mathrm{st}})=\left\vert -\right\rangle \left\langle -\right\vert
\sum_{R}p_{R}^{\mathrm{st}}(t)|R).  \label{MphotonConfo}
\end{equation}%
Here, the weights $\{p_{R}^{\mathrm{st}}(t)\}$ satisfy the normalization $%
\sum_{R}p_{R}^{\mathrm{st}}(t)=1,$ and are defined as 
\begin{equation}
p_{R}^{\mathrm{st}}(t)\equiv \frac{\gamma _{R}\left\langle +\right\vert \rho
_{R}^{\mathrm{st}}(t)\left\vert +\right\rangle }{\sum_{R^{\prime }}\gamma
_{R^{\prime }}\left\langle +\right\vert \rho _{R^{\prime }}^{\mathrm{st}%
}(t)\left\vert +\right\rangle },  \label{PrStConfo}
\end{equation}%
where the notation $\rho _{R}^{\mathrm{st}}(t)=(R|\rho _{t}^{\mathrm{st}})$
was used. From Eqs.~(\ref{SystemConfiSTOCH}) and (\ref{MphotonConfo}), it is
simple to get%
\begin{equation}
\rho _{S}^{\mathrm{st}}(t)\rightarrow (1|\mathcal{\hat{M}}_{\mathrm{ph}%
}|\rho _{t}^{\mathrm{st}})=\left\vert -\right\rangle \left\langle
-\right\vert ,  \label{RhoSChange}
\end{equation}%
and that%
\begin{equation}
(R|P_{t}^{\mathrm{st}})\rightarrow \mathrm{Tr}_{S}[(R|\mathcal{\hat{M}}_{%
\mathrm{ph}}|\rho _{t}^{\mathrm{st}})]=p_{R}^{\mathrm{st}}(t).
\label{PrChange}
\end{equation}%
Eq.~(\ref{RhoSChange}) shows that in fact, after a photon detection event
the system collapses to its lower state $\left\vert -\right\rangle .$ On the
other hand, Eq.~(\ref{PrChange}) say us that $p_{R}^{\mathrm{st}}(t)$ is the
value of the configurational populations after a photon recording event.

\begin{figure}[tb]
\includegraphics[bb=26 15 380 560,angle=0,width=7.5 cm]{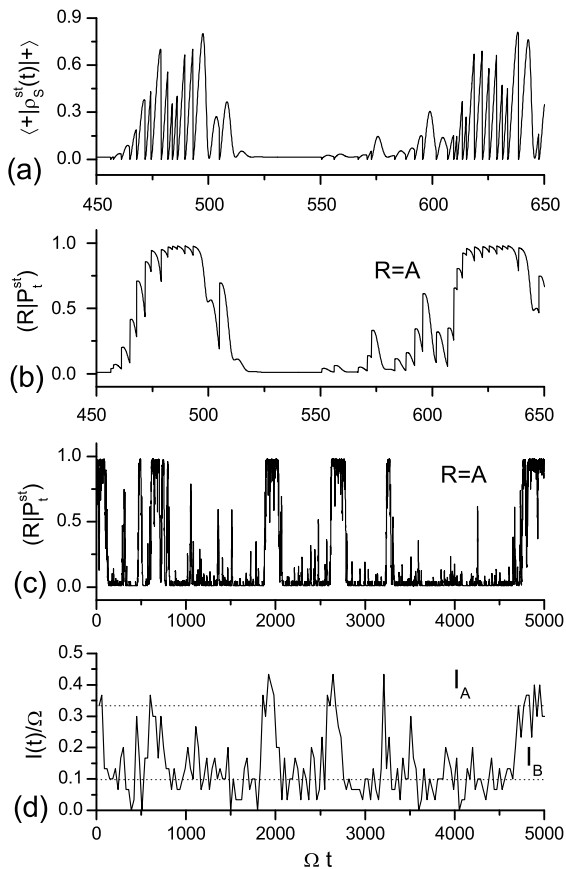}
\caption{Stochastic realizations of a fluorophore system defined by the
evolution Eq.~(\protect\ref{LindbladSelf}). The configurational space is
two-dimensional, $R=A,B.$ The parameters are $\Omega _{R}=\Omega ,$ $\protect%
\delta \protect\omega _{R}=0,$ $\protect\gamma _{A}/\Omega =1,$ $\protect%
\gamma _{B}/\Omega =10,$ $\protect\phi _{AB}/\Omega =0.003,$ $\protect\phi %
_{BA}/\Omega =0.009.$ The laser is in resonance with the system, i.e., $%
\protect\omega _{L}=\protect\omega _{0}.$ (a) Realization of the of the
upper population of the system $\left\langle +\right\vert \protect\rho _{S}^{%
\mathrm{st}}(t)\left\vert +\right\rangle .$ (b)-(c) Realization of the
configurational population of the bath $(R|P_{t}^{\mathrm{st}}),$ for $R=A.$
(d) Intensity realization. The values $I_{R}$ are defined by Eq.~(\protect
\ref{IntR}).}
\label{Figura1_SMS}
\end{figure}
Between the detection events the stochastic dynamics is defined the
conditional evolution defined by the superoperator $\mathcal{\hat{D}}$
[Eqs.~(\ref{DConfo}) and (\ref{RhoCondConfo})]. On the other hand, as there
exist only one measurement apparatus, the weights $\{\mathrm{t}_{\mu }(t)\},$
Eq.~(\ref{pu}), here reduce to $\mathrm{t}_{\mathrm{ph}}(t)=1.$

In\ the next figures, we consider a fluorophore system coupled to an
environment characterized by a two-dimensional configurational space, $%
R=A,B, $ which only affect the decay rates $\{\gamma _{R}\}$ of the system,
i.e., the Rabi frequencies [Eq.~(\ref{Hamiltonian})] do not depend on the
configurational states, $\Omega _{R}=\Omega ,$ and the spectral shifts [Eq.~(%
\ref{SpectralShifts})] are null, $\delta \omega _{R}=0.$ Furthermore, the
laser is in resonance with the system, i.e., $\omega _{L}=\omega _{0}.$

In Fig.~1(a) we show a realization of the upper population of the system $%
\left\langle +\right\vert \rho _{S}^{\mathrm{st}}(t)\left\vert
+\right\rangle $ [Eq.~(\ref{SystemConfiSTOCH})]. The realizations were
determined by using the finite time step algorithm defined in Appendix D.
Each collapse of the upper population to zero is related to a photon
emission [see Eq.~(\ref{RhoSChange})].

In Fig.~1(b), we show the realization of the configurational population $%
(R|P_{t}^{\mathrm{st}})$ for $R=A$ [Eq.~(\ref{SystemConfiSTOCH})]. As the
configurational space is two-dimensional, $(A|P_{t}^{\mathrm{st}})+(B|P_{t}^{%
\mathrm{st}})=1.$ We remark that these realizations are associated to a
measurement process that only gives information about the photon emission
events. Not any information is provided about the configurational states of
the bath. Therefore, the realizations of $(R|P_{t}^{\mathrm{st}})$ are the
best estimation\ \cite{tsang} about the configurational state of the
reservoir that can be obtained by knowing the master equation (\ref%
{LindbladSelf}) and a given realization of the photon detector apparatus.

In Fig 1(c) we plot $(R|P_{t}^{\mathrm{st}})$ (for\ $R=A$) over a larger
time interval. For the chosen parameter values, the configurational
populations develop a quasi-dichotomic behavior. When $(R|P_{t}^{\mathrm{st}%
})\approx 1,$ we can affirm that is highly probable that the bath is in the
configurational state $|R).$

In Fig.~1(d), we plot the scattered intensity, which is defined by $%
I(t)=[n(t+\delta t)-n(t)]/\delta t,$ where $n(t)$ is the number of photon
recording events up to time $t$ and $\delta t$ is an adequate time flag
averaging. Its fluctuations are highly correlated with the values of $%
(R|P_{t}^{\mathrm{st}}).$ In fact, the intensity fluctuates around two
well-defined values $I_{R},$ which are defined by the intensity of a
Markovian fluorescent system \cite{breuerbook,plenio,carmichaelbook}
characterized by the parameters corresponding to each configurational state 
\cite{OpenSMS}, i.e.,%
\begin{equation}
I_{R}=\frac{\gamma _{R}\Omega _{R}^{2}}{\gamma _{R}^{2}+2\Omega
_{R}^{2}+4\delta _{R}^{2}},  \label{IntR}
\end{equation}%
where $\delta _{R}\equiv \omega _{L}-\omega _{R}.$ The dichotomic behavior
arises because the system is able to emit a large number of photons
previously to the occurrence of a transition between the configurational
bath states, i.e., $\sum_{R^{\prime }}\phi _{R^{\prime }R}\ll I_{R}.$

\begin{figure}[tb]
\includegraphics[bb=35 20 405 565,angle=0,width=7.5 cm]{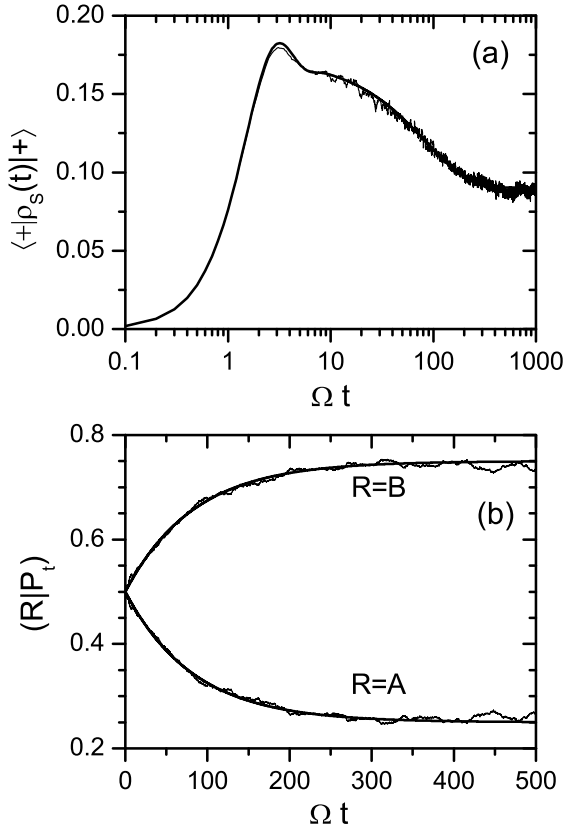}
\caption{Evolution of the upper population $\left\langle +\right\vert 
\protect\rho _{S}(t)\left\vert +\right\rangle $ (a) and the configurational
populations $(R|P_{t})$ (b), that follows from Eqs.~(\protect\ref%
{LindbladSelf}) and (\protect\ref{ClassicalPconfo}) respectively. The
parameters are the same than in Figure 1. The initial conditions are $%
\protect\rho _{S}(0)=\left\vert -\right\rangle \left\langle -\right\vert $
and $(A|P_{0})=(B|P_{0})=1/2.$ The noisy curves follow from an average over
the realizations shown in Fig. 1.}
\label{Figura2_SMS}
\end{figure}

In Fig.~2(a) we plot the upper population $\left\langle +\right\vert \rho
_{S}(t)\left\vert +\right\rangle $ that follows from Eq.~(\ref{LindbladSelf}%
), as well as an average over ($\approx $10$^{3}$) realizations of $%
\left\langle +\right\vert \rho _{S}^{\mathrm{st}}(t)\left\vert
+\right\rangle $ [see Fig.~1(a)]. In Fig.~2(b), we plot the analytical
solution of the configurational populations $(R|P_{t})$\ defined by Eq.~(\ref%
{ClassicalPconfo}), as well as an average over realization of $(R|P_{t}^{%
\mathrm{st}})$ [see Fig.~1(b) and (c)]. In both cases the ensemble averages
recover the dynamics dictated by the corresponding master equations, showing
the consistency of the developed approach.

\subsubsection{Photon emission process}

The recording events are characterized by the stochastic waiting time
distribution Eq.~(\ref{Wst_mu}). Then, we write%
\begin{equation}
w_{\mathrm{st}}(t,t^{\prime })=\mathrm{Tr}_{S}[(1|\mathcal{\hat{J}}e^{%
\mathcal{\hat{D}}(t-t^{\prime })}\mathcal{\hat{M}}_{\mathrm{ph}}|\rho
_{t^{\prime }}^{\mathrm{st}})].  \label{Westocastica}
\end{equation}%
Notice that here $\mathcal{\hat{J}}=\mathcal{\hat{J}}_{\mathrm{ph}}.$ The
function $w_{\mathrm{st}}(t,t^{\prime })$ define the statistic of the time
intervals between consecutive photon emissions. From Eqs.~(\ref{MphotonSelf}%
) and (\ref{MphotonConfo}) it follows 
\begin{equation}
w_{\mathrm{st}}(t,t^{\prime })=\sum_{RR^{\prime }}\gamma _{R}\left\langle
+\right\vert e_{RR^{\prime }}^{\mathcal{\hat{D}}(t-t^{\prime })}[\left\vert
-\right\rangle \left\langle -\right\vert p_{R^{\prime }}^{\mathrm{st}%
}(t^{\prime })]\left\vert +\right\rangle .  \label{wstConfo}
\end{equation}%
This expression allows us to get an analytical expression for $w_{\mathrm{st}%
}(t,t^{\prime })$ [not provided due to its extension] that parametrically
depends on the set $\{p_{R^{\prime }}^{\mathrm{st}}(t^{\prime })\},$ Eq.~(%
\ref{PrStConfo}).

\begin{figure}[tb]
\includegraphics[bb=35 20 405 565,angle=0,width=7.5 cm]{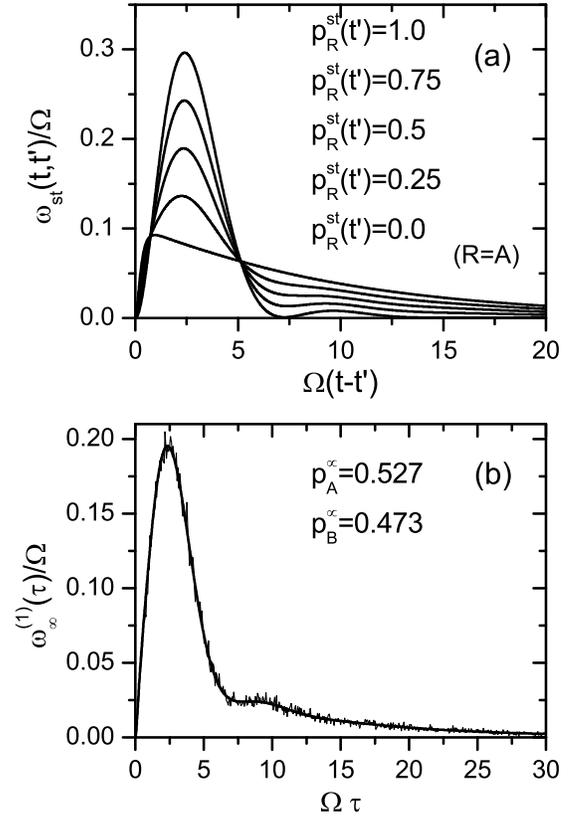}
\caption{(a) Plot of the waiting time distribution $w_{\mathrm{st}%
}(t,t^{\prime })$ [Eq.~(\protect\ref{wstConfo})] for different values of $%
p_{R}^{\mathrm{st}}(t^{\prime }).$ From top to bottom we take $(R=A)$ $%
p_{R}^{\mathrm{st}}(t^{\prime })=1,$ $0.75,$ $0.5,$ $0.25$ and $0.$ (b)
Stationary waiting time distribution [Eq.~(\protect\ref{WEstacionConfo})].
The stationary weights [Eq.~(\protect\ref{p_r_Estacion})] read $%
p_{A}^{\infty }=0.527$ and $p_{B}^{\infty }=0.473.$ The noisy curve
corresponds to a numerical distribution determine from the intervals between
consecutive photon emissions [Eq.~(\protect\ref{WaitUnoEstacion})] along a
single realization (like that shown in Fig. 1). In both plots, the
parameters are the same than in Fig. 1.}
\label{Figura3_SMS}
\end{figure}

As the set of weights $\{p_{R^{\prime }}^{\mathrm{st}}(t^{\prime })\}$
correspond to the configurational populations after a photon recording event
[see Eq.~(\ref{PrChange})], the waiting time distribution change between
consecutive photon emissions. The successive (stochastic) values of $%
\{p_{R^{\prime }}^{\mathrm{st}}(t^{\prime })\}$ can be read from the
realization of $(R|P_{t}^{\mathrm{st}})$ shown in Fig. 1(b). Notice that in
each event, defined by the collapses $\left\langle +\right\vert \rho _{S}^{%
\mathrm{st}}(t)\left\vert +\right\rangle \rightarrow 0$ [Fig. 1(a)], $%
(R|P_{t}^{\mathrm{st}})$ suffer an abrupt change in its slope.

In Fig.~3(a), we plot $w_{\mathrm{st}}(t,t^{\prime })$ (as a function of $%
t-t^{\prime }$) for different values of the parameters $p_{R^{\prime }}^{%
\mathrm{st}}(t^{\prime }).$ As the configurational space is two-dimensional,
the two parameters satisfy the normalization $p_{A}^{\mathrm{st}}(t^{\prime
})+p_{B}^{\mathrm{st}}(t^{\prime })=1.$ We notice that $w_{\mathrm{st}%
}(t,t^{\prime })$ has a strong dependence on the values of the
configurational populations $\{p_{R^{\prime }}^{\mathrm{st}}(t^{\prime })\},$
which in turn say us that the photon emission process strongly departs from
a renewal one. For Markovian fluorescent systems, the set $\{p_{R^{\prime
}}^{\mathrm{st}}(t^{\prime })\}$ reduce to only one parameter with value
equal to one (the configurational space is one-dimensional). Therefore, $w_{%
\mathrm{st}}(t,t^{\prime })$ is the same object along a measurement
trajectory, recovering a renewal process.

\subsubsection{Stationary waiting time distributions}

\begin{figure}[tb]
\includegraphics[bb=35 20 405 565,angle=0,width=7.5 cm]{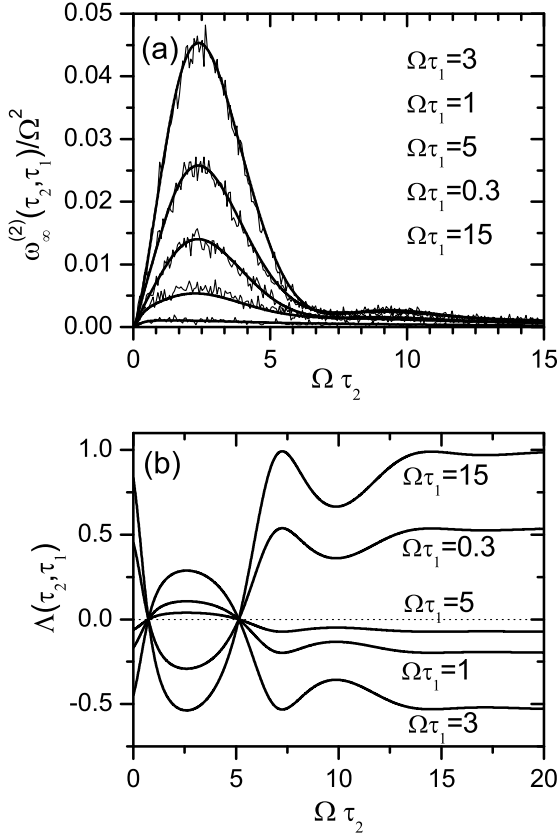}
\caption{(a) Stationary two-time waiting time distribution $w_{\infty
}^{(2)}(\protect\tau _{2},\protect\tau _{1})$ [Eq.~(\protect\ref%
{WTwoEstacionConfo})] for different values of $\protect\tau _{1}.$ From top
to bottom we take $\Omega \protect\tau _{1}=3,$ $1,$ $5,$ $0.3,$ and $15.$
The noisy curves correspond to the time average Eq.~(\protect\ref%
{WaitDosEstacion}). (b) Parameter $\Lambda (\protect\tau _{2},\protect\tau %
_{1})$ [Eq.~(\protect\ref{Lambda})] determine from Eqs.~(\protect\ref%
{WEstacionConfo}) and (\protect\ref{WTwoEstacionConfo}), for different
values of $\protect\tau _{1}.$ From top to bottom $\Omega \protect\tau %
_{1}=15,$ $0.3,$ $5,$ $1,$ and $3.$ The parameters are the same than in Fig.
1.}
\label{Figura4_SMS}
\end{figure}

By measuring the time intervals between successive photon emissions along a
given trajectory one can determine the stationary waiting time distribution
Eq.~(\ref{WaitUnoEstacion}). The analytical expression for this probability
distribution can be read from Eq.~(\ref{waitST1}). We get%
\begin{equation}
w_{\infty }^{(1)}(\tau )=\sum_{RR^{\prime }}\gamma _{R}\left\langle
+\right\vert e_{RR^{\prime }}^{\mathcal{\hat{D}}\tau }[\left\vert
-\right\rangle \left\langle -\right\vert p_{R^{\prime }}^{\infty
}]\left\vert +\right\rangle ,  \label{WEstacionConfo}
\end{equation}%
where the weights $p_{R}^{\infty }$ are defined from the relation%
\begin{equation}
\mathcal{\hat{M}}_{\mathrm{ph}}|\rho _{\infty })=\left\vert -\right\rangle
\left\langle -\right\vert \sum_{R}p_{R}^{\infty }|R),  \label{M_Estacion}
\end{equation}%
delivering the expression%
\begin{equation}
p_{R}^{\infty }\equiv \frac{\gamma _{R}\left\langle +\right\vert \rho
_{R}^{\infty }\left\vert +\right\rangle }{\sum_{R^{\prime }}\gamma
_{R^{\prime }}\left\langle +\right\vert \rho _{R^{\prime }}^{\infty
}\left\vert +\right\rangle }.  \label{p_r_Estacion}
\end{equation}%
Here, $\rho _{R}^{\infty }\equiv \lim_{t\rightarrow \infty }\rho
_{R}(t)=(R|\rho _{\infty })$ [Eq.~(\ref{RhoInfinity})]. By comparing Eq.~(%
\ref{WEstacionConfo}) with Eq.~(\ref{wstConfo}), we realize that $w_{\infty
}^{(1)}(\tau )$ follows from $w_{\mathrm{st}}(t,t^{\prime })$ after the
replacements $p_{R^{\prime }}^{\mathrm{st}}(t^{\prime })\rightarrow
p_{R^{\prime }}^{\infty }$ and $(t-t^{\prime })\rightarrow \tau .$

In Fig.~3(b) we plot the analytical expression for $w_{\infty }^{(1)}(\tau )$
that follows from Eq.~(\ref{WEstacionConfo}). Furthermore, we show a
numerical distribution determine from the time average Eq.~(\ref%
{WaitUnoEstacion}). The theoretical distribution correctly fit the numerical
result. Consistently, we also checked that the time average of $\{p_{R}^{%
\mathrm{st}}(t^{\prime })\}$ along a single trajectory recover the weights $%
\{p_{R}^{\infty }\},$ Eq.~(\ref{p_r_Estacion}). 
\begin{figure}[tb]
\includegraphics[bb=0 0 250 265,angle=0,width=6.75 cm]{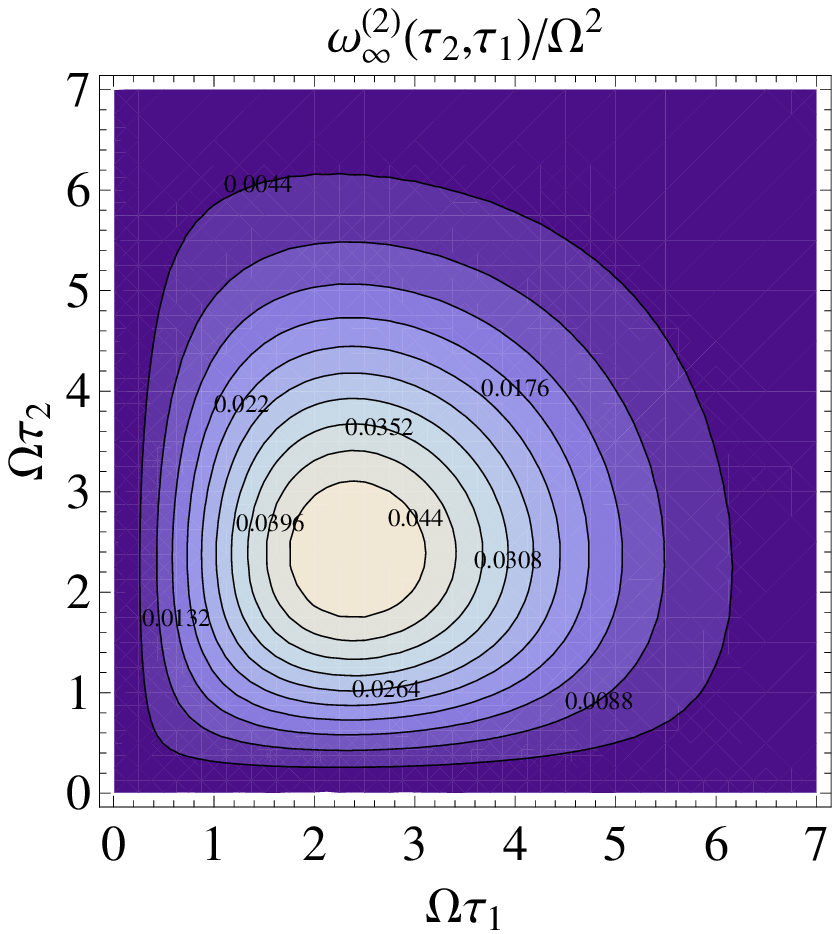}
\includegraphics[bb=0 0 250 265,angle=0,width=6.75 cm]{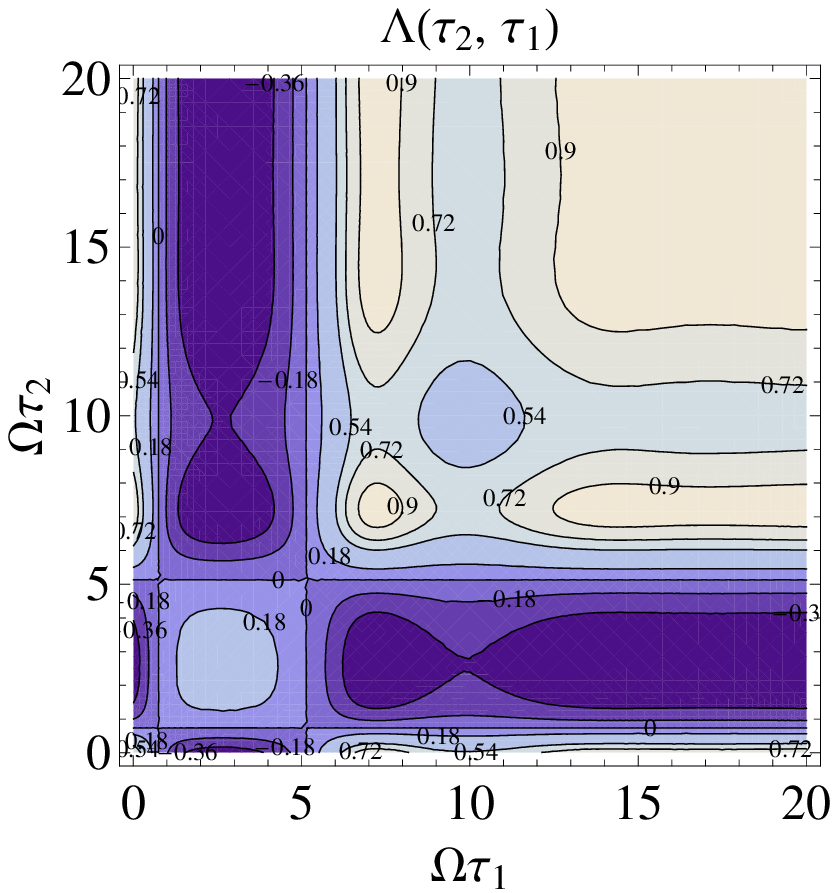}
\caption{Contour plot of $w_{\infty }^{(2)}(\protect\tau _{2},\protect\tau %
_{1})/\Omega ^{2}$ [upper panel] and $\Lambda (\protect\tau _{2},\protect%
\tau _{1})$ [lower panel] corresponding to the plots of Fig. 4.}
\label{Figura5_SMS}
\end{figure}

The analytical expression for the second waiting time Eq.~(\ref%
{WaitDosEstacion}) can be obtained from Eq.~(\ref{waitST2}), delivering%
\begin{equation}
w_{\infty }^{(2)}(\tau _{2},\tau _{1})=\sum_{RR^{\prime }}\gamma
_{R}\left\langle +\right\vert e_{RR^{\prime }}^{\mathcal{\hat{D}}\tau
_{2}}[\left\vert -\right\rangle \left\langle -\right\vert \varphi
_{R^{\prime }}(\tau _{1})]\left\vert +\right\rangle ,
\label{WTwoEstacionConfo}
\end{equation}%
where the functions $\varphi _{R}(\tau _{1})$ read%
\begin{equation}
\varphi _{R}(\tau _{1})\equiv \gamma _{R}\sum_{R^{\prime }}\left\langle
+\right\vert e_{RR^{\prime }}^{\mathcal{\hat{D}}\tau _{1}}[\left\vert
-\right\rangle \left\langle -\right\vert p_{R^{\prime }}^{\infty
}]\left\vert +\right\rangle .
\end{equation}

In Fig.~4(a) we plot the analytical expression for $w_{\infty }^{(2)}(\tau
_{2},\tau _{1})$ that follows from Eq.~(\ref{WTwoEstacionConfo}) for
different values of $\tau _{1}$. Furthermore, we show the numerical result
that follows by determining the probability distribution of two consecutive
time intervals between successive photon emissions along a single
trajectory, Eq.~(\ref{WaitDosEstacion}). The theoretical result correctly
fit the numerical distribution.

In Fig.~4(b) we plot the dimensionless parameter $\Lambda (\tau _{2},\tau
_{1})$ [Eq.~(\ref{Lambda})] determine from Eqs.~(\ref{WEstacionConfo}) and (%
\ref{WTwoEstacionConfo}), for different values of $\tau _{1}.$ For almost
all values of $\tau _{2}$ and $\tau _{1},$ $\Lambda (\tau _{2},\tau _{1})$
departs appreciably from zero, indicating the departure of the photon
emission process from a renewal one. We also note that there exist special
values of the consecutive time intervals where $\Lambda (\tau _{2},\tau
_{1})=0.$ From the definition Eq.~(\ref{Lambda}), we deduce that when $%
\Lambda (\tau _{2},\tau _{1})>0,$ the frequency of the successive intervals $%
\tau _{1}$ and $\tau _{2}$ is greater than in the renewal case. The
situation $\Lambda (\tau _{2},\tau _{1})<0,$ admits the inverse
interpretation. For clarifying the structure of both $w_{\infty }^{(2)}(\tau
_{2},\tau _{1})$ and $\Lambda (\tau _{2},\tau _{1}),$ in Fig.~5 we show
their contour plots. As can be deduced from Fig.~4 and 5, $\Lambda (\tau
_{2},\tau _{1})$ reach it maximal values for higher values of both $\tau _{2}
$ and $\tau _{1}.$

\subsubsection{Slow and fast environment fluctuations}

The expressions for the stochastic waiting distribution $w_{\mathrm{st}%
}(t,t^{\prime })$ [Eq.~(\ref{wstConfo})], and the first [$w_{\infty
}^{(1)}(\tau ),$ Eq.~(\ref{WEstacionConfo})] and second [$w_{\infty
}^{(2)}(\tau _{2},\tau _{1}),$ Eq.~(\ref{WTwoEstacionConfo})] stationary
waiting time distributions allow to characterize the photon emission process
as well as its departure with respect to a renewal one. Here, we provide
simple analytical expressions for these objects in the limit of both fast
and slow environment fluctuations.

The characteristic time of the bath fluctuations are measured by the rates $%
\{\phi _{R^{\prime }R}\},$ Eqs.~(\ref{LindbladSelf}) and (\ref%
{ClassicalPconfo}). On the other hand, the average time between photon
emissions is measured by the inverse of the intensities $\{I_{R}\},$ Eq.~(%
\ref{IntR}).

When the bath fluctuations are much slower than the average time between
photon emissions, $\{\phi _{R^{\prime }R}\}\ll \{I_{R}\},$ it is valid to
approximate the conditional evolution defined by the superoperator $\mathcal{%
\hat{D}}$\ [Eqs.~(\ref{DConfo}) and (\ref{UnNormalized})] as%
\begin{equation}
e_{RR^{\prime }}^{\mathcal{\hat{D}}(t-t^{\prime })}[\left\vert
-\right\rangle \left\langle -\right\vert p_{R^{\prime }}^{\mathrm{st}%
}(t^{\prime })]\approx \delta _{RR^{\prime }}e_{RR}^{\mathcal{\hat{D}}%
(t-t^{\prime })}[\left\vert -\right\rangle \left\langle -\right\vert p_{R}^{%
\mathrm{st}}(t^{\prime })].  \label{Aproximator}
\end{equation}%
This approximation corresponds to disregarding the non-diagonal
contributions between photon recording events. By inserting this condition
in Eq.~(\ref{wstConfo}) we get%
\begin{eqnarray}
w_{\mathrm{st}}(t,t^{\prime }) &\simeq &\sum_{R}\gamma _{R}\left\langle
+\right\vert e_{RR}^{\mathcal{\hat{D}}(t-t^{\prime })}[\left\vert
-\right\rangle \left\langle -\right\vert p_{R}^{\mathrm{st}}(t^{\prime
})]\left\vert +\right\rangle ,\ \ \ \   \notag \\
&=&\sum_{R}w_{R}(t-t^{\prime })p_{R}^{\mathrm{st}}(t^{\prime }).
\label{WEstoAproxiConfo}
\end{eqnarray}%
Then, $w_{\mathrm{st}}(t,t^{\prime })$ can be written as a linear
combination of the waiting time distributions $\{w_{R}(t)\},$ each one being
defined by the expression%
\begin{equation}
w_{R}(t)\equiv \gamma _{R}\left\langle +\right\vert e_{RR}^{\mathcal{\hat{D}}%
t}[\left\vert -\right\rangle \left\langle -\right\vert ]\left\vert
+\right\rangle .  \label{WaitingMarkovTiempo}
\end{equation}%
This function correspond to the waiting time distribution associated to a
Markovian fluorescent system \cite{carmichaelbook,carmichael} with decay
rate $\gamma _{R},$ and whose Hamiltonian is given by $H_{R},$ Eq.~(\ref%
{Hamiltonian}), i.e., its transition frequency is $\omega _{R}=(\omega
_{0}+\delta \omega _{R}),$ and its coupling to the external laser is
measured by $\Omega _{R}.$ This result can straightforwardly be read from
Eqs.~(\ref{UnNormalized}) and (\ref{RhoCondConfo}) under the replacement $%
\phi _{RR^{\prime }}\rightarrow 0.$ In the Laplace domain, $t\rightarrow u,$
it can be written as \cite{rapid}%
\begin{equation}
w_{R}(u)=\frac{\gamma _{R}/2}{u+\gamma _{R}/2}\left( \frac{\Omega
_{R}^{2}h_{R}(u)}{u^{2}+u\gamma _{R}+\Omega _{R}^{2}h_{R}(u)}\right) ,
\label{WrMarkov}
\end{equation}%
where the auxiliary function $h_{R}(u)$\ is%
\begin{equation}
h_{R}(u)=\frac{(u+\gamma _{R}/2)^{2}}{(u+\gamma _{R}/2)^{2}+\delta _{R}^{2}},
\end{equation}%
and $\delta _{R}=\omega _{L}-\omega _{R}.$ After Laplace inversion, we get%
\begin{equation}
w_{R}(t)=\frac{2\gamma _{R}\Omega _{R}^{2}}{\zeta _{R}}\exp (-\gamma
_{R}t/2)[\cosh (\xi _{R}^{+}t)-\cosh (\xi _{R}^{-}t)],
\label{WrMarkovTiempo}
\end{equation}%
where $\xi _{R}^{\pm }=[\gamma _{R}^{2}-4(\Omega _{R}^{2}+\delta
_{R}^{2})\pm \zeta _{R}]^{1/2}/(2\sqrt{2}),$ with $\zeta _{R}=\{[\gamma
_{R}^{2}+4(\Omega _{R}^{2}+\delta _{R}^{2})]^{2}-16\gamma _{R}^{2}\Omega
_{R}^{2}\}^{1/2}.$ When $\delta _{R}=0,$ the expression of Refs.~\cite%
{carmichaelbook,carmichael} is recovered.

We have checked that Eq.~(\ref{WEstoAproxiConfo}) joint with Eq.~(\ref%
{WrMarkovTiempo}) provide an excellent approximation to the exact functions
plotted in Fig.~3(a). On the other hand, Eq.~(\ref{Aproximator}) is also
useful for approximating the stationary waiting time distributions. Eq.~(\ref%
{WEstacionConfo}) leads to%
\begin{equation}
w_{\infty }^{(1)}(\tau )\simeq \sum_{R}w_{R}(\tau )p_{R}^{\infty },
\label{WUnoAproximadaConfo}
\end{equation}%
while from Eq.~(\ref{WTwoEstacionConfo}), we get%
\begin{equation}
w_{\infty }^{(2)}(\tau _{2},\tau _{1})\simeq \sum_{R}w_{R}(\tau
_{2})w_{R}(\tau _{1})p_{R}^{\infty }.  \label{WDOSAproximadaConfo}
\end{equation}%
Furthermore, under the hypothesis of slow fluctuations, from Eq.~(\ref%
{LindbladSelf}) we can approximate $\gamma _{R}\left\langle +\right\vert
\rho _{R}^{\infty }\left\vert +\right\rangle \simeq I_{R}P_{R}^{\infty }.$
The constants $I_{R}$ [Eq.~(\ref{IntR})] are the intensities associated to
each configurational state $R.$ On the other hand, $P_{R}^{\infty }$ are the
stationary values of the configurational populations Eq.~(\ref%
{ClassicalPconfo}), i.e., $P_{R}^{\infty }\equiv \lim_{t\rightarrow \infty
}P_{R}(t).$ Therefore, from Eq.~(\ref{p_r_Estacion}) we get the approximate
expression%
\begin{equation}
p_{R}^{\infty }\simeq \frac{I_{R}P_{R}^{\infty }}{\sum_{R^{\prime
}}I_{R^{\prime }}P_{R^{\prime }}^{\infty }}.  \label{prAproximadoConfo}
\end{equation}

Eqs.~(\ref{WUnoAproximadaConfo}), (\ref{WDOSAproximadaConfo}), and (\ref%
{prAproximadoConfo}), also provide an excellent approximation to the exact
analytical results plotted in Figs.~3, 4 and 5. Moreover, they have a clear
physical meaning. In the slow limit each bath state establishes an intensity
regime defined by Eq.~(\ref{IntR}). Hence, the statistic of the non-renewal
photon emission process follows from an average of the renewal statistic
associated to each state [defined by $w_{R}(\tau )$]. The weight of each
contribution is $p_{R}^{\infty }.$ Consistently, this factors, which are the
average configurational populations after a detection event [Eq. (\ref%
{M_Estacion})], are proportional to the intensities $I_{R}$\ and the
stationary populations $P_{R}^{\infty }$\ related to each bath state.

When the bath fluctuations are much faster than the average time between
photon emissions, $\{\phi _{R^{\prime }R}\}\gg \{I_{R}\},$ the
configurational populations reach their stationary values, $P_{R}^{\infty
}=\lim_{t\rightarrow \infty }P_{R}(t),$ before happening many photon
emissions. Hence, the fluorophore behaves as a Markovian fluorescent system
with decay rate $\bar{\gamma}\equiv \sum_{R}\gamma _{R}P_{R}^{\infty },$
Rabi frequency $\bar{\Omega}\equiv \sum_{R}\Omega _{R}P_{R}^{\infty },$ and
detuning $\bar{\delta}\equiv \sum_{R}\delta _{R}P_{R}^{\infty }.$ The photon
emission process becomes a renewal one, being defined by the waiting time
distribution Eq.~(\ref{WrMarkovTiempo}) with $\{\gamma _{R},\Omega
_{R},\delta _{R}\}\rightarrow \{\bar{\gamma},\bar{\Omega},\bar{\delta}\}.$
Near of this limit, for two-dimensional configurational spaces, $w_{\infty
}^{(2)}(\tau _{2},\tau _{1})$ develops small asymmetries on its arguments $%
w_{\infty }^{(2)}(\tau _{2},\tau _{1})\neq w_{\infty }^{(2)}(\tau _{1},\tau
_{2}).$ In general, this property may arises in the intermediate regime
between fast, $w_{\infty }^{(2)}(\tau _{2},\tau _{1})\simeq w_{\infty
}^{(1)}(\tau _{2})w_{\infty }^{(1)}(\tau _{1}),$ and slow bath fluctuations,
Eq.~(\ref{WDOSAproximadaConfo}).

The configurational fluctuations are frozen when Eq.~(\ref{LindbladSelf}) is
defined with $\{\phi _{R^{\prime }R}\}=0.$ This case was partially addressed
in Refs. \cite{rapid,JPB}. Our present treatment provides a general
description. Evidently, the configurational populations remain unaffected
during all the evolution, $|P_{t})=|P_{0}).$ The results of Ref.~\cite%
{rapid,JPB} follows from the approximation $|P_{t}^{\mathrm{st}})\approx
|P_{0}),$ which is valid in a weak laser intensity regime and when the
dynamics develops two different times scales induced by an infinite
dimensional configurational space.

\subsection{Light Assisted environment fluctuations}

The general evolution Eq.~(\ref{LindbladRate}) may also cover the case in
which the statistical properties of the radiation pattern, as well as the
environment fluctuations, depend on the external laser intensity \cite%
{OpenSMS,luzAssisted}, i.e., light assisted processes. By taking $A=\sigma ,$
and $\eta _{RR^{\prime }}\rightarrow \gamma _{RR^{\prime }},$ we write%
\begin{eqnarray}
\dfrac{d\rho _{R}(t)}{dt}\!\! &=&\!\!\dfrac{-i}{\hbar }[H_{R},\rho
_{R}(t)]\!-\!\gamma _{R}(\{D,\rho _{R}(t)\}\!_{+}-\!\mathcal{J}[\rho
_{R}(t)])  \notag \\
&&\!\!\!\!-\!\sum\limits_{R^{\prime }}\!\gamma _{R^{\prime }\!R}\{D,\rho
_{R}(t)\}\!_{+}+\!\sum\limits_{R^{\prime }}\!\gamma _{RR^{\prime }}\mathcal{J%
}[\rho _{R^{\prime }}(t)],  \label{LigthMaster}
\end{eqnarray}%
where $D$ and $\mathcal{J}$ follows from Eq.~(\ref{DandJ}). In this case,
the evolution of the configurational populations [Eq.~(\ref{Prob})] strongly
depend on the state of the system. In fact, here the configurational
transitions may only occur when a photon emission happens. Thus, in general
it is not possible to write a simple equation for their evolution. Only when 
$\{\gamma _{RR^{\prime }}\}\ll \{\gamma _{R}\},$ a classical rate equation
similar to Eq.~(\ref{ClassicalPconfo}) can be derived \cite%
{OpenSMS,luzAssisted}.

\subsubsection{Photon measurement operator}

Here, the photon-measurement superoperator $\mathcal{\hat{M}}_{\mathrm{ph}%
}|\rho )=\mathcal{\hat{J}}_{\mathrm{ph}}|\rho )/\mathrm{Tr}_{S}(1|\mathcal{%
\hat{J}}_{\mathrm{ph}}|\rho ),$ from Eq.~(\ref{LigthMaster}), reads%
\begin{equation}
\mathcal{\hat{M}}_{\mathrm{ph}}|\rho )=\frac{\sum_{R}|R)\{\gamma _{R}\sigma
\rho _{R}\sigma ^{\dagger }+\sum_{R^{\prime }}\gamma _{RR^{\prime }}\sigma
\rho _{R^{\prime }}\sigma ^{\dagger }\}}{\sum_{R^{\prime \prime }}\tilde{%
\gamma}_{R^{\prime \prime }}\mathrm{Tr}_{S}[\sigma ^{\dagger }\sigma \rho
_{R^{\prime \prime }}]},  \label{MphotonLigth}
\end{equation}%
where $|\rho )=\sum_{R}|R)\rho _{R},$ and we have defined the rate%
\begin{equation}
\tilde{\gamma}_{R}\equiv \gamma _{R}+\sum\nolimits_{R^{\prime }}\gamma
_{R^{\prime }R}.  \label{tildeGamma}
\end{equation}%
As in the previous case [Eq.~(\ref{MphotonSelf})], Eq.~(\ref{MphotonLigth})
take into account all possible configurational paths that lead to a photon
emission. The conditional evolution defined by the operator $\mathcal{\hat{D}%
}=\mathcal{\hat{L}}-\mathcal{\hat{J}}_{\mathrm{ph}},$ expressed through the
evolution of the unnormalized state $|\rho _{t}^{\mathrm{u}})$ [Eq.~(\ref%
{UnNormalized})] reads%
\begin{equation}
\dfrac{d\rho _{R}^{\mathrm{u}}(t)}{dt}=\dfrac{-i}{\hbar }[H_{R},\rho _{R}^{%
\mathrm{u}}(t)]-\tilde{\gamma}_{R}\{D,\rho _{R}^{\mathrm{u}}(t)\}_{+}.
\label{Dligth}
\end{equation}%
Notice that in contrast with Eq.~(\ref{RhoCondConfo}), here the conditional
evolution is diagonal in the $R$-space.

\subsubsection{Stochastic dynamics}

The structure of the stochastic dynamics of the vectorial state $|\rho _{t}^{%
\mathrm{st}}),$ Eq.~(\ref{stochastic}), is similar to that of the previous
case. When a photon detection event happens it implies the transformation%
\begin{equation}
|\rho _{t}^{\mathrm{st}})\rightarrow \mathcal{\hat{M}}_{\mathrm{ph}}|\rho
_{t}^{\mathrm{st}})=\left\vert -\right\rangle \left\langle -\right\vert
\sum_{R}p_{R}^{\mathrm{st}}(t)|R),  \label{JumpEnLuz}
\end{equation}%
where the weights satisfies $\sum_{R}p_{R}^{\mathrm{st}}=1.$ From Eq.~(\ref%
{MphotonLigth}), here they read%
\begin{equation}
p_{R}^{\mathrm{st}}(t)\equiv \frac{\gamma _{R}\left\langle +\right\vert \rho
_{R}^{\mathrm{st}}(t)\left\vert +\right\rangle +\sum_{R^{\prime }}\gamma
_{RR^{\prime }}\left\langle +\right\vert \rho _{R^{\prime }}^{\mathrm{st}%
}(t)\left\vert +\right\rangle }{\sum_{R^{\prime \prime }}\tilde{\gamma}%
_{R^{\prime \prime }}\left\langle +\right\vert \rho _{R^{\prime \prime }}^{%
\mathrm{st}}(t)\left\vert +\right\rangle }.  \label{prLucecita}
\end{equation}%
From Eq.~(\ref{SystemConfiSTOCH}) it is simple to demonstrate that Eqs.~(\ref%
{RhoSChange}) and (\ref{PrChange}) are also valid in this case. Therefore,
in each photon recording event the system collapse to its ground state while 
$p_{R}^{\mathrm{st}}(t)$ define the posterior value of the configurational
populations. 
\begin{figure}[tb]
\includegraphics[bb=26 15 380 560,angle=0,width=7.5 cm]{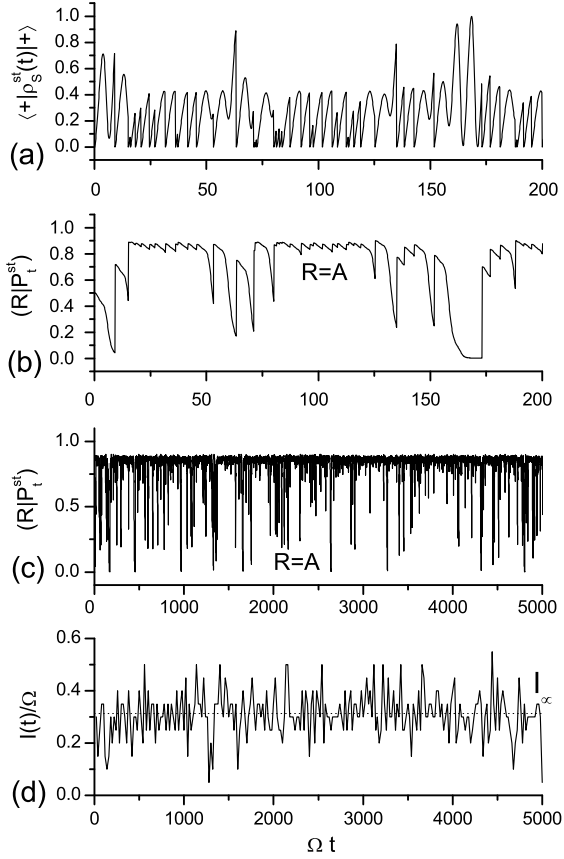}
\caption{Stochastic realizations of a fluorophore system defined by the
evolution Eq.~(\protect\ref{LigthMaster}). The configurational space is
two-dimensional, $R=A,B.$ The parameters are $\Omega _{R}=\Omega ,$ $\protect%
\delta \protect\omega _{R}=0,$ $\protect\gamma _{A}/\Omega =1.8,$ $\protect%
\gamma _{B}/\Omega =0.15,$ $\protect\gamma _{AB}/\Omega =0.35,$ $\protect%
\gamma _{BA}/\Omega =0.2.$ The laser is in resonance with the system, i.e., $%
\protect\omega _{L}=\protect\omega _{0}.$ (a) Realization of the of the
upper population of the system $\left\langle +\right\vert \protect\rho _{S}^{%
\mathrm{st}}(t)\left\vert +\right\rangle .$ (b)-(c) Realization of the
configurational population of the bath $(R|P_{t}^{\mathrm{st}}),$ for $R=A.$
(d) Intensity realization. $I_{\infty }$ is defined by Eq.~(\protect\ref%
{IntInfy}).}
\label{Figura6_SMS}
\end{figure}

In\ the next figures we consider a fluorophore system whose evolution is
defined by Eq.~(\ref{LigthMaster}) and a two-dimensional configurational
space, $R=A,B.$ The Rabi frequencies [Eq.~(\ref{Hamiltonian})] do not depend
on the configurational states, $\Omega _{R}=\Omega ,$ and the spectral
shifts [Eq.~(\ref{SpectralShifts})] are null, $\delta \omega _{R}=0.$
Therefore, the bath states only affect the decay rates $\{\gamma _{R}\}$ of
the system. The laser is in resonance with the system, i.e., $\omega
_{L}=\omega _{0}.$

In Fig.~6(a) we show a realization of the upper population of the system $%
\left\langle +\right\vert \rho _{S}^{\mathrm{st}}(t)\left\vert
+\right\rangle $ [Eq.~(\ref{SystemConfiSTOCH})]. The times of the photon
emission events correspond to the collapse of the upper population to zero.
Fig.~6(b) shows the realization of the configurational population of the
bath $(R|P_{t}^{\mathrm{st}}),$ for $R=A$ [Eq.~(\ref{SystemConfiSTOCH})].
Due to the chosen parameter values, at any time it is not possible to
predict with total certainty $[(R|P_{t}^{\mathrm{st}})\approx 1]$ the
configurational state of the reservoir. This fact is evident from Fig.~6(c),
where we plot $(R|P_{t}^{\mathrm{st}})$ (for\ $R=A$) over a larger time
interval. In Fig.~6(d), we plot the scattered intensity. Consistently with
the behavior of $(R|P_{t}^{\mathrm{st}}),$ the intensity does not develop
any dichotomic behavior. The intensity fluctuates around the value $%
I_{\infty }$ defined by (see Eq.~(55) in Ref.~\cite{OpenSMS})%
\begin{equation}
I_{\infty }=\sum_{R}\tilde{\gamma}_{R}\left\langle +\right\vert \rho
_{R}^{\infty }\left\vert +\right\rangle ,  \label{IntInfy}
\end{equation}%
where as before $\rho _{R}^{\infty }=\lim_{t\rightarrow \infty }\rho
_{R}(t). $

In Fig.~7(a) and (b) we plot the analytical solutions of the upper
population $\left\langle +\right\vert \rho _{S}(t)\left\vert +\right\rangle $
and the configurational populations $(R|P_{t})$\ that follows from Eq.~(\ref%
{LigthMaster}) [and Eq.~(\ref{Prob})]. The noisy curves correspond to an
average over realizations like those shown in Fig.~6. Notice that here, the
behavior of the configurational population strongly depart from an
exponential one, indicating that their underlying dynamics is highly
non-Markovian. The physical origin of this characteristic is the dependence
of the configurational transitions on the system state.

\begin{figure}[tb]
\includegraphics[bb=35 20 405 565,angle=0,width=7.5 cm]{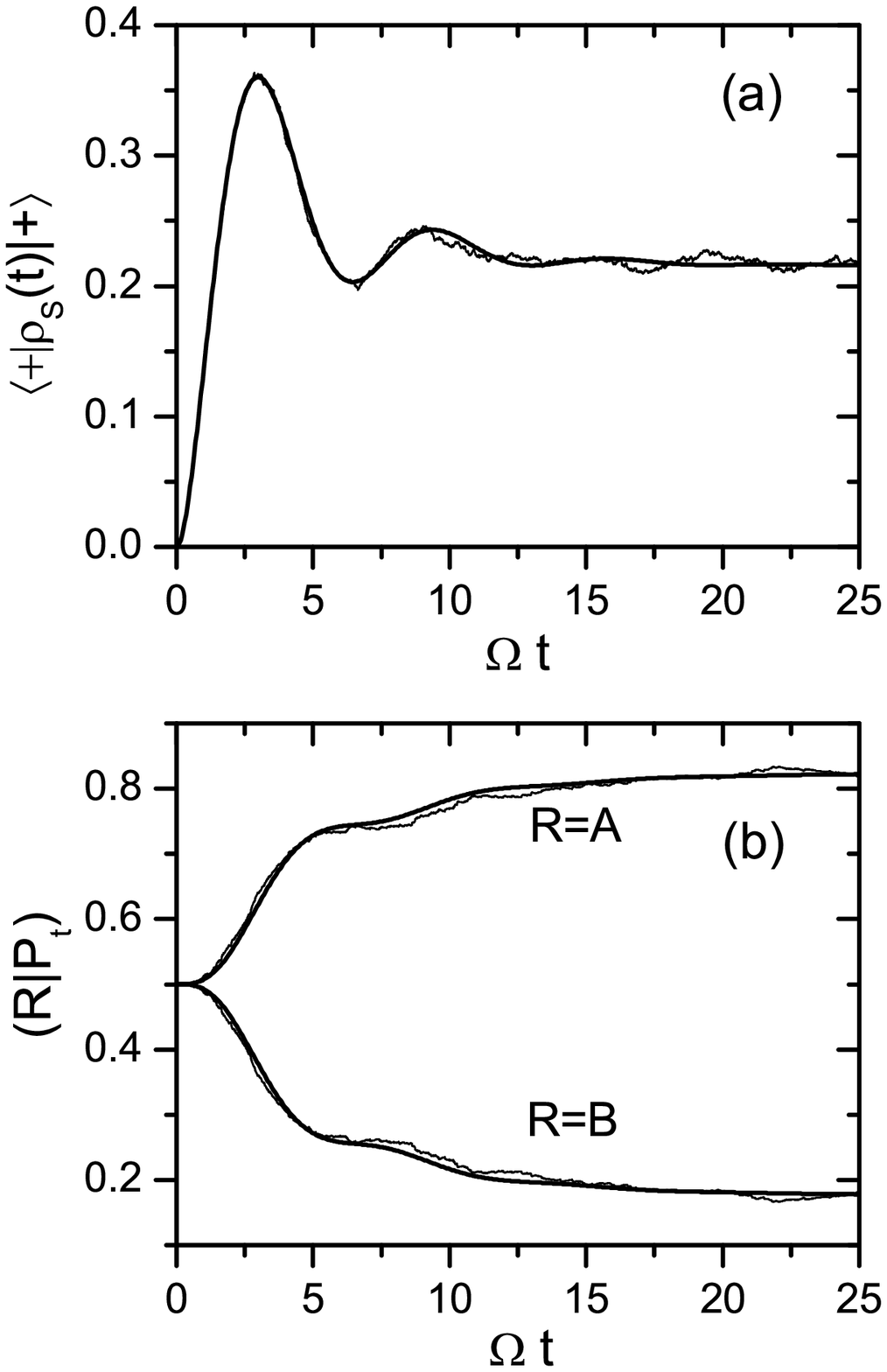}
\caption{Evolution of the upper population $\left\langle +\right\vert 
\protect\rho _{S}(t)\left\vert +\right\rangle $ (a) and the configurational
populations $(R|P_{t})=\mathrm{Tr}_{S}[(R\left\vert \protect\rho _{t}\right)
]$ (b) that follow from Eq.~(\protect\ref{LigthMaster}). The parameters are
the same than in Fig. 6. The initial conditions are $\protect\rho %
_{S}(0)=\left\vert -\right\rangle \left\langle -\right\vert $ and $%
(A|P_{0})=(B|P_{0})=1/2.$ The noisy curves follow from an average over the
realizations shown in Fig. 6.}
\label{Figura7_SMS}
\end{figure}

\subsubsection{Photon emission process}

In this case it is also possible to define a stochastic waiting time
distribution that parametrically depends on the configurational populations
after a photon detection event, i.e., $w_{\mathrm{st}}(t,t^{\prime })=%
\mathrm{Tr}_{S}[(1|\mathcal{\hat{J}}e^{\mathcal{\hat{D}}(t-t^{\prime })}%
\mathcal{\hat{M}}_{\mathrm{ph}}|\rho _{t^{\prime }}^{\mathrm{st}})],$ Eq.~(%
\ref{Westocastica}). From Eqs.~(\ref{MphotonLigth}) and (\ref{JumpEnLuz}) we
get 
\begin{equation}
w_{\mathrm{st}}(t,t^{\prime })=\sum_{RR^{\prime }}\tilde{\gamma}%
_{R}\left\langle +\right\vert e_{RR^{\prime }}^{\mathcal{\hat{D}}%
(t-t^{\prime })}[\left\vert -\right\rangle \left\langle -\right\vert
p_{R^{\prime }}^{\mathrm{st}}(t^{\prime })]\left\vert +\right\rangle ,
\end{equation}%
where $p_{R^{\prime }}^{\mathrm{st}}(t^{\prime })$ is given by Eq.~(\ref%
{prLucecita}). As the vectorial superoperator $\mathcal{\hat{D}}$ is
diagonal [see Eq.~(\ref{Dligth})], this expression can be written as%
\begin{equation}
w_{\mathrm{st}}(t,t^{\prime })=\sum_{R}\tilde{w}_{R}(t-t^{\prime })p_{R}^{%
\mathrm{st}}(t^{\prime }),  \label{WestocasticaLigth}
\end{equation}%
where $\tilde{w}_{R}(t)$ is defined by Eq.~(\ref{WrMarkovTiempo}) after the
replacement $\gamma _{R}\rightarrow \tilde{\gamma}_{R}$ [Eq.~(\ref%
{tildeGamma})]. While Eq.~(\ref{WEstoAproxiConfo}) is an approximation valid
in the limit of slow environmental fluctuations, here Eq.~(\ref%
{WestocasticaLigth}) is valid independently of the values of any of the
parameters that define the system evolution, Eq.~(\ref{LigthMaster}).

In Fig.~8(a), we plot $w_{\mathrm{st}}(t,t^{\prime })$ (as a function of $%
t-t^{\prime }$) for different values of the parameters $p_{R^{\prime }}^{%
\mathrm{st}}(t^{\prime }).$ As in the previous case, $w_{\mathrm{st}%
}(t,t^{\prime })$ has a strong dependence on the values of the
configurational populations $\{p_{R^{\prime }}^{\mathrm{st}}(t^{\prime })\},$
implying strong departures from a renewal process. In fact, notice that
depending on $p_{R^{\prime }}^{\mathrm{st}}(t^{\prime }),$ $w_{\mathrm{st}%
}(t,t^{\prime })$ may or not to develops oscillatory behaviors.

\begin{figure}[tb]
\includegraphics[bb=35 20 405 565,angle=0,width=7.5 cm]{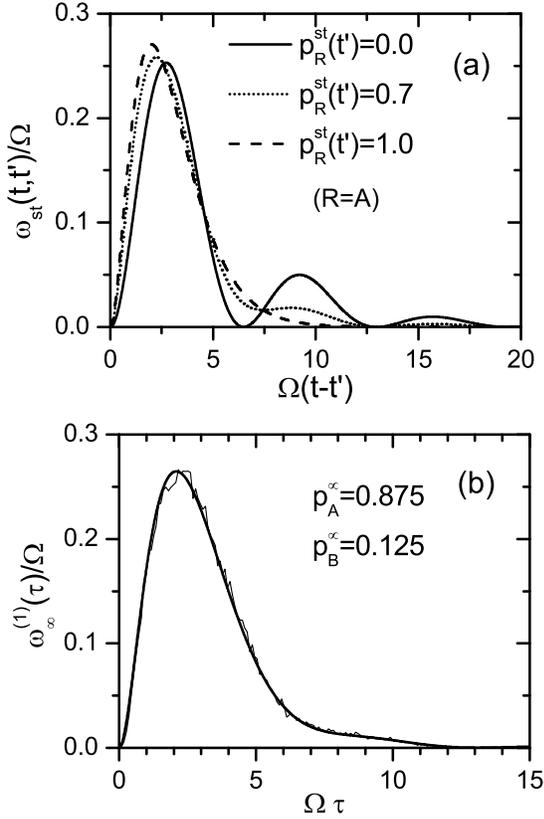}
\caption{(a) Plot of the waiting time distribution $w_{\mathrm{st}%
}(t,t^{\prime })$ [Eq.~(\protect\ref{WestocasticaLigth})] for different
values of $p_{R}^{\mathrm{st}}(t^{\prime }).$ We take the values $p_{R}^{%
\mathrm{st}}(t^{\prime })=0$ (full line), $0.7$ (dotted line), and $1$
(dashed line), where $R=A.$ (b) Stationary waiting time distribution [Eq.~(%
\protect\ref{WUnoEstacionariaLigth})]. The stationary weights [Eq.~(\protect
\ref{PesosEstacionarioLigth})] read $p_{A}^{\infty }=0.875$ and $%
p_{B}^{\infty }=0.125.$ The noisy curve corresponds to the time average Eq.~(%
\protect\ref{WaitUnoEstacion}). The parameters are the same than in Fig.~6.}
\label{Figura8_SMS}
\end{figure}

\subsubsection{Stationary waiting time distributions}

The first stationary waiting time distribution Eq.~(\ref{WaitUnoEstacion})
from Eq.~(\ref{waitST1}) can be written as $w_{\infty }^{(1)}(\tau
)=\sum_{RR^{\prime }}\tilde{\gamma}_{R}\left\langle +\right\vert
e_{RR^{\prime }}^{\mathcal{\hat{D}}\tau }[\left\vert -\right\rangle
\left\langle -\right\vert p_{R}^{\infty }]\left\vert +\right\rangle .$ After
using the definition of the conditional evolution Eq.~(\ref{Dligth}), it
follows%
\begin{equation}
w_{\infty }^{(1)}(\tau )=\sum_{R}\tilde{w}_{R}(\tau )p_{R}^{\infty },
\label{WUnoEstacionariaLigth}
\end{equation}%
where the weights $p_{R}^{\infty }$ are determine from the relation $%
\mathcal{\hat{M}}_{\mathrm{ph}}|\rho _{\infty })=\left\vert -\right\rangle
\left\langle -\right\vert \sum_{R}p_{R}^{\infty }|R),$ delivering%
\begin{equation}
p_{R}^{\infty }\equiv \frac{\gamma _{R}\left\langle +\right\vert \rho
_{R}^{\infty }\left\vert +\right\rangle +\sum_{R^{\prime }}\gamma
_{RR^{\prime }}\left\langle +\right\vert \rho _{R^{\prime }}^{\infty
}\left\vert +\right\rangle }{\sum_{R^{\prime \prime }}\tilde{\gamma}%
_{R^{\prime \prime }}\left\langle +\right\vert \rho _{R^{\prime \prime
}}^{\infty }\left\vert +\right\rangle }.  \label{PesosEstacionarioLigth}
\end{equation}

In Fig.~8(b) we plot the analytical expression for $w_{\infty }^{(1)}(\tau )$
[Eq.~(\ref{WUnoEstacionariaLigth})] joint with the numerical distribution
(noisy curve) obtained as the probability distribution of the time intervals
between successive photon emissions along a single trajectory, Eq.~(\ref%
{WaitUnoEstacion}). The theoretical and numerical results match between them.

The second waiting time distribution $w_{\infty }^{(2)}(\tau _{2},\tau _{1})$
follows from Eq.~(\ref{waitST2}). After some calculations we get%
\begin{eqnarray}
w_{\infty }^{(2)}(\tau _{2},\tau _{1}) &=&\sum_{R}\left\{ \tilde{w}_{R}(\tau
_{2})q_{R}+\sum_{R^{\prime }}\tilde{w}_{R^{\prime }}(\tau _{2})q_{R^{\prime
}R}\right\}  \notag \\
&&\times \tilde{w}_{R}(\tau _{1})p_{R}^{\infty }.
\label{WDOSEstacionariaLigth}
\end{eqnarray}%
Here, we introduced the factors%
\begin{equation}
q_{R}\equiv \frac{\gamma _{R}}{\tilde{\gamma}_{R}},\ \ \ \ \ \ \ \ \ \ \ \ \
\ \ q_{R^{\prime }R}\equiv \frac{\gamma _{R^{\prime }R}}{\tilde{\gamma}_{R}},
\label{q_erres}
\end{equation}%
which for any $R$ satisfy the normalization $q_{R}+\sum\nolimits_{R^{\prime
}}q_{R^{\prime }R}=1.$

The physical content of Eq.~(\ref{WDOSEstacionariaLigth}) can be read as
follows. After a first photon emission [contribution $\tilde{w}_{R}(\tau
_{1})p_{R}^{\infty }],$ the second one happens without a configurational
transition with probability $q_{R},$ while with probability $q_{R^{\prime
}R} $ it is endowed with the configurational transition $R\rightarrow
R^{\prime }.$ This interpretation is consistent with the results presented
in Ref.~\cite{luzAssisted} (see also Appendix A). On the other hand, while
in general $w_{\infty }^{(2)}(\tau _{2},\tau _{1})\neq w_{\infty
}^{(2)}(\tau _{1},\tau _{2}),$ here for two-dimensional configurational
spaces $w_{\infty }^{(2)}(\tau _{2},\tau _{1})$ is symmetric on its
arguments.

\begin{figure}[tb]
\includegraphics[bb=35 20 405 565,angle=0,width=7.5 cm]{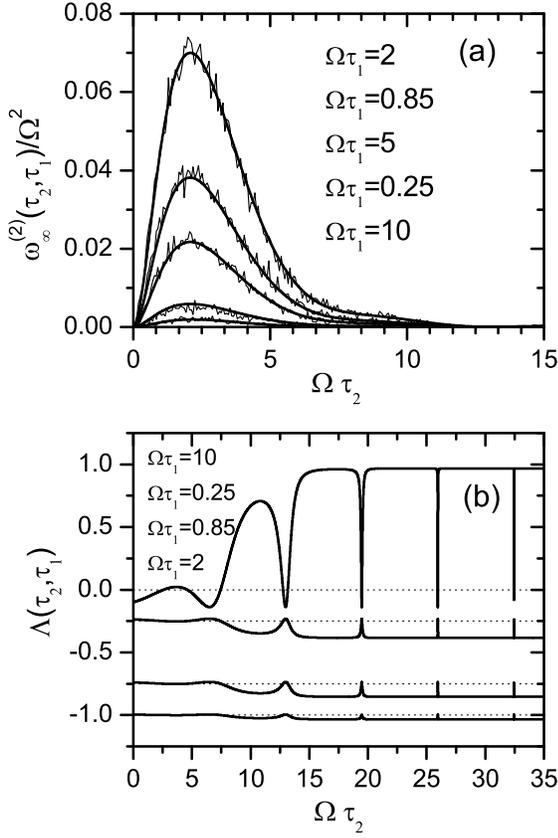}
\caption{(a) Stationary two-time waiting time distribution $w_{\infty
}^{(2)}(\protect\tau _{2},\protect\tau _{1})$ [Eq.~(\protect\ref%
{WDOSEstacionariaLigth})] for different values of $\protect\tau _{1}.$ From
top to bottom we take $\Omega \protect\tau _{1}=2,$ $0.85,$ $5,$ $0.25,$ and 
$10.$ The noisy curves correspond to the time average Eq.~(\protect\ref%
{WaitDosEstacion}). (b) Parameter $\Lambda (\protect\tau _{2},\protect\tau %
_{1})$ [Eq.~(\protect\ref{Lambda})] determine from Eqs.~(\protect\ref%
{WUnoEstacionariaLigth}) and (\protect\ref{WDOSEstacionariaLigth}), for
different values of $\protect\tau _{1}.$ From top to bottom $\Omega \protect%
\tau _{1}=10,$ $0.25,$ $0.85,$ and $2.$ For clarity, the curves
corresponding to the last three values where shifted by $-0.25,$ $-0.75,$
and $-1$ respectively. The parameters are the same than in Fig.~6.}
\label{Figura9_SMS}
\end{figure}
\begin{figure}[tb]
\includegraphics[bb=0 0 250 265,angle=0,width=6.75cm]{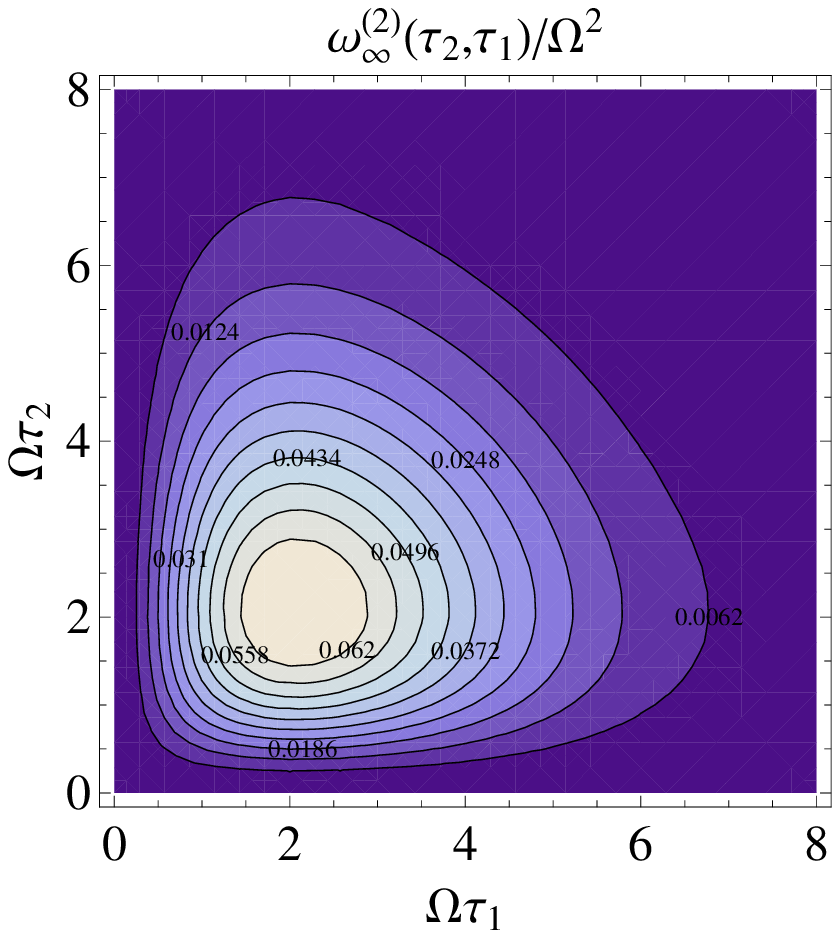}
\includegraphics[bb=0 0 250 265,angle=0,width=6.75cm]{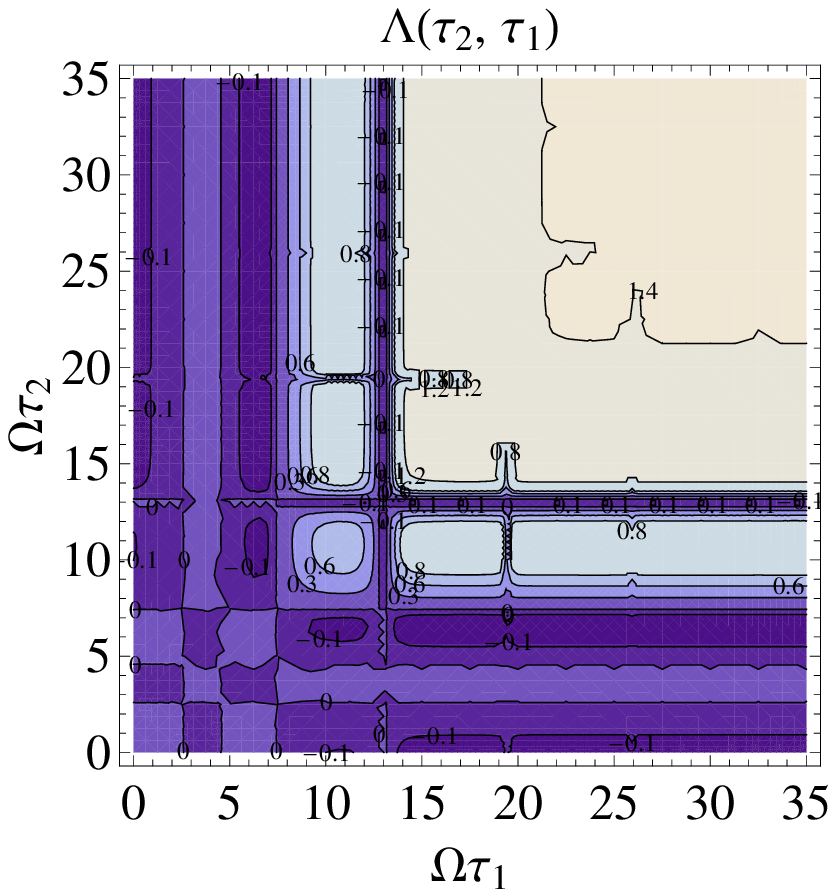}
\caption{Contour plot of $w_{\infty }^{(2)}(\protect\tau _{2},\protect\tau %
_{1})/\Omega ^{2}$ [upper panel] and $\Lambda (\protect\tau _{2},\protect%
\tau _{1})$ [lower panel] corresponding to the plots of Fig. 9.}
\label{Figura10_SMS}
\end{figure}

In Fig.~9(a) we plot $w_{\infty }^{(2)}(\tau _{2},\tau _{1})$ [Eq.~(\ref%
{WDOSEstacionariaLigth})] for different values of $\tau _{1}$. We also show
the numerical distribution, Eq.~(\ref{WaitDosEstacion}). In Fig.~9(b) we
plot the dimensionless parameter $\Lambda (\tau _{2},\tau _{1})$ [Eq.~(\ref%
{Lambda})] determine from Eqs.~(\ref{WUnoEstacionariaLigth}) and (\ref%
{WDOSEstacionariaLigth}), for different values of $\tau _{1}.$ In Fig.~10 we
show its contour plot as well as that corresponding to $w_{\infty
}^{(2)}(\tau _{2},\tau _{1}).$ For small values of of $\tau _{2}$ and $\tau
_{1},$ $\Lambda (\tau _{2},\tau _{1})$ is almost null, while for higher
values of both times $\Lambda (\tau _{2},\tau _{1})$ reaches its maximal
values.

We notice that the structure of $w_{\infty }^{(2)}(\tau _{2},\tau _{1})$ is
very similar to that shown in Fig.~\ref{Figura5_SMS}. The same affirmation
is valid for the corresponding $w_{\infty }^{(1)}(\tau ),$ i.e., Fig.~\ref%
{Figura3_SMS}(b) and \ref{Figura8_SMS}(b). Nevertheless, here $\Lambda (\tau
_{2},\tau _{1})$ [Fig.~\ref{Figura10_SMS}] develops a much richer structure
or dependence in $\tau _{2}$ and $\tau _{1}.$ The origin of this
characteristic can be related with the behavior of the underlying
photon-to-photon emission process. In fact, in the light assisted case $w_{%
\mathrm{st}}(t,t^{\prime })$ [Fig.~\ref{Figura8_SMS}(a)], depending on the
values of the parameter $p_{R}^{\mathrm{st}}(t^{\prime }),$ may develops
strong oscillatory behaviors, while in Fig.~\ref{Figura3_SMS}(a) the
behaviors are almost monotonous. Independently of the underlying
environmental dynamic, by increasing the external laser intensity the
renewal departure measure $\Lambda (\tau _{2},\tau _{1})$ develops a richer
structure.

When the radiation pattern develops a blinking phenomenon \cite{OpenSMS},
i.e., for slow (light-assisted) environment fluctuations, $\{\gamma
_{RR^{\prime }}\}\ll \{\gamma _{R}\},$ Eq.~(\ref{PesosEstacionarioLigth})
becomes 
\begin{equation}
p_{R}^{\infty }\simeq \frac{\tilde{I}_{R}\tilde{P}_{R}^{\infty }}{%
\sum_{R^{\prime }}\tilde{I}_{R^{\prime }}\tilde{P}_{R^{\prime }}^{\infty }},
\end{equation}%
where $\tilde{I}_{R}$ follows from Eq.~(\ref{IntR}) under the replacement $%
\gamma _{R}\rightarrow \tilde{\gamma}_{R}$ [Eq.~(\ref{tildeGamma})], and the
probabilities $\tilde{P}_{R}^{\infty }$ are the stationary solution of a
classical master equation obtained from Eq.~(\ref{ClassicalPconfo}) under
the replacement $\phi _{RR^{\prime }}\rightarrow \Gamma _{R^{\prime
}R}=q_{R^{\prime }R}\tilde{I}_{R}$ (see Eqs.~(81) and (82) in Ref.~\cite%
{OpenSMS}). In the limit of fast environment fluctuations, $\{\gamma
_{RR^{\prime }}\}\gg \{\gamma _{R}\},$ the photon emission process becomes a
renewal one.

\section{Summary and Conclusions}

In this paper, we formulated a quantum-jump approach for describing the
radiation patterns of single fluorescent systems coupled to complex
fluctuating environments. Our results rely on a density matrix formulation
of the problem. The master Eq.~(\ref{LindbladRate}) take into account both
the system dynamic as well as a width class of environment fluctuations.

The quantum-jump approach relies on a quantum measurement theory. Here,
after introducing general measurement transformations acting on the system
and the configurational bath space [Eq.~(\ref{measurement})], the density
matrix evolution was written as an average over measurement trajectories,
Eq.~(\ref{Unravelling}). The weight of each trajectory is measured by its
associated $n$-joint probability, Eq.~(\ref{Joint}). The hierarchy of these
objects completely characterizes the statistical properties of the
measurement processes. Its functional form in an asymptotic time regime
provides information about observables defined from a time average along a
single measurement trajectory. Eqs.~(\ref{waitST1}) define the stationary
probability distribution for the time interval between consecutive
measurement events, while Eq.~(\ref{waitST2}) define the joint probability
for two consecutive intervals. These two objects allow measuring the
departure of the measurement process from a renewal one.

The decomposition of the density matrix evolution into a set of measurement
trajectories leads to a stochastic representation of the system dynamics,
Eq.~(\ref{stochastic}). Each stochastic realization can be related with a
particular measurement trajectory. Their structure allowed us to define how
the measurement process occurs event-to-event. The waiting time distribution
Eq.~(\ref{Wst_mu}) defines the probability density for consecutive recording
events. In contrast with a renewal process, it depends on the stochastic
state of the system, property that breaks the renewal character of the
measurement process. This dependence encodes the influence of the bath
fluctuations.

The case when there exist only one measurement process, providing
information about the photon emission events, was analyzed in detail.
Independently of the underlying bath dynamics (and the measurement
processes, see Appendix A) the photon-to-photon emission process is defined
by a stochastic waiting time distribution that parametrically depends on the
configurational bath populations. The analysis based on Eq.~(\ref%
{LindbladSelf}) allows to describing situations like spectral diffusion
process, lifetime fluctuations and molecules diffusing in a solution. The
stochastic waiting time distribution, Eq.~(\ref{wstConfo}), and the first
and second stationary waiting time distributions, Eqs.~(\ref{WEstacionConfo}%
) and (\ref{WTwoEstacionConfo}) respectively, provides a deep
characterization of the photon emission process. These general expressions
assume a simple form when the environment fluctuations are much slower than
the optical system transitions. In fact, in such a case those objects can be
written as linear combinations of the waiting time distribution associated
to a Markovian fluorescent system characterized by the parameters
corresponding to each configurational state, Eqs.~(\ref{WEstoAproxiConfo}), (%
\ref{WUnoAproximadaConfo}), and (\ref{WDOSAproximadaConfo}). The case of
light assisted process, Eq.~(\ref{LigthMaster}), admits a similar
description, Eqs.~(\ref{WestocasticaLigth}), (\ref{WUnoEstacionariaLigth}),
and (\ref{WDOSEstacionariaLigth}).

The developed results provide an alternative theoretical tool for analyzing
single fluorescent systems coupled to classically fluctuating environments.
In fact, the explicit analytical characterization of statistical observables
like the stationary waiting time distributions, Eqs.~(\ref{WaitUnoEstacion})
and (\ref{WaitDosEstacion}), and the renewal departure function, Eq.~(\ref%
{Lambda}), may provide a power tool for deducing the underlying structure of
complex nanoscopic reservoirs analyzed through fluorescence spectroscopy.

\section*{Acknowledgments}

The author thanks fruitful discussions with R. Rebolledo, F. Petruccione, M.
Orszag, and A. Barchielli at the \textquotedblleft 30th Conference on
Quantum Probability and Related Topics,\textquotedblright\ (2009) Santiago,
Chile. This work was supported by CONICET, Argentina.

\appendix

\section{Measuring photon emissions and configurational transitions}

In Section IV we characterized the quantum jump approach (for both
self-environment fluctuations and light assisted processes) when the
measurement action only gives information about the photon emission events.
While that is the standard situation in SMS, the formalism developed in
Section III allow us to analyze the case in which there exist extra
measurement channels (apparatus) that give information about the
configurational states of the reservoir. Besides its theoretical interest
and potential applications, the following analysis also allows to understand
some previous results \cite{luzAssisted}.

Here, we assume that at any time one know which is the configurational state
of the bath. Under this condition, the stochastic dynamic of $|\rho _{t}^{%
\mathrm{st}})$ and $|P_{t}^{\mathrm{st}})$ assume the structure%
\begin{equation}
|\rho _{t}^{\mathrm{st}})=\rho _{S}^{\mathrm{st}}(t)|R_{t}^{\mathrm{st}}),\
\ \ \ \ \ \ \ \ \ \ |P_{t}^{\mathrm{st}})=|R_{t}^{\mathrm{st}}),
\label{single}
\end{equation}%
where $\mathrm{Tr}_{S}[\rho _{S}^{\mathrm{st}}(t)]=1,$ and $R_{t}^{\mathrm{st%
}}$ randomly change over the set of possible values $R=1,2,\cdots R_{\max }.$
Therefore, here the vectorial nature of $|\rho _{t}^{\mathrm{st}})$ can be
avoided. In fact, all relevant information is encoded in $\rho _{S}^{\mathrm{%
st}}(t)$ and $(R|P_{t}^{\mathrm{st}})=\delta _{RR_{t}^{\mathrm{st}}}$ [see
Eq.~(\ref{SystemConfiSTOCH})]. While the underlying master equations are
different, the results of Ref. \cite{petruccione} also rely on the previous
assumption.

\subsection{Self-fluctuating environments}

First we analyze the case of self-fluctuating environments, Eq.~(\ref%
{LindbladSelf}). The parameter $\mu $ includes one term corresponding to the
photon detector, $\mu =\mathrm{ph},$ and $\mu =1\cdots R_{\max }$ terms that
detect (measure) when a transition to a given conformational state $R$
happens.

\subsubsection{Measurement operators}

The measurement operators [Eq.~(\ref{measurement})] read 
\begin{subequations}
\label{MesureSelectiveConfoGENERAL}
\begin{eqnarray}
\mathcal{\hat{M}}_{\mathrm{ph}}|\rho ) &=&\frac{\sum_{R}\gamma _{R}|R)\
\sigma \rho _{R}\sigma ^{\dagger }}{\sum_{R^{\prime }}\gamma _{R^{\prime }}\ 
\mathrm{Tr}_{S}[\sigma ^{\dagger }\sigma \rho _{R^{\prime }}]}, \\
\mathcal{\hat{M}}_{R}|\rho ) &=&\frac{|R)\sum_{R^{\prime }}\phi _{RR^{\prime
}}\rho _{R^{\prime }}}{\sum_{R^{\prime \prime }}\phi _{RR^{\prime \prime }}%
\mathrm{Tr}_{S}[\rho _{R^{\prime \prime }}]},
\end{eqnarray}%
where $|\rho )=\sum_{R}|R)\rho _{R}.$ The (unnormalized) conditional
evolution [Eq.~(\ref{UnNormalized})] is diagonal in the $R$-space and reads 
\end{subequations}
\begin{equation}
\dfrac{d\rho _{R}^{\mathrm{u}}(t)}{dt}=\dfrac{-i}{\hbar }[H_{R},\rho _{R}^{%
\mathrm{u}}(t)]-\gamma _{R}\{D,\rho _{R}^{\mathrm{u}}(t)\}_{+}-\tilde{\phi}%
_{R}\rho _{R}^{\mathrm{u}}(t),  \label{CondicionalPhotonU}
\end{equation}%
where the rate $\tilde{\phi}_{R}$ is defined by%
\begin{equation}
\tilde{\phi}_{R}\equiv \sum\nolimits_{R^{\prime }}\phi _{R^{\prime }R}.
\label{PhiNormalizadaR}
\end{equation}%
These definitions provide a splitting of Eq.~(\ref{LindbladSelf}) that
allows to formulate the quantum-jump approach, Eq.~(\ref{vectorial}). $%
\mathcal{\hat{M}}_{\mathrm{ph}}$ corresponds to the transformation
associated to a photon detection event. On the other hand, $\mathcal{\hat{M}}%
_{R}$ take in account all transitions $R^{\prime }\rightarrow R$ that leave
the bath in the configurational state $R.$

\subsubsection{Stochastic dynamics}

The measurement operators Eq.~(\ref{MesureSelectiveConfoGENERAL}) imply the
transformations $[|\rho _{t}^{\mathrm{st}})=\rho _{S}^{\mathrm{st}%
}(t)|R_{t}^{\mathrm{st}})]$%
\begin{subequations}
\label{MesureSelectiveConfo}
\begin{eqnarray}
|\rho _{t}^{\mathrm{st}}) &\rightarrow &\mathcal{\hat{M}}_{\mathrm{ph}}|\rho
_{t}^{\mathrm{st}})=\left\vert -\right\rangle \left\langle -\right\vert
|R_{t}^{\mathrm{st}}),  \label{colapso} \\
|\rho _{t}^{\mathrm{st}}) &\rightarrow &\mathcal{\hat{M}}_{R}|\rho _{t}^{%
\mathrm{st}})=\rho _{S}^{\mathrm{st}}(t)|R).  \label{transicion}
\end{eqnarray}
The first transformation collapse the system to its ground state and does
not affect the configurational state. The measurement operator $\mathcal{%
\hat{M}}_{R}$ leaves invariant the system state $\rho _{S}^{\mathrm{st}}(t),$
while produces the configurational transition $|R_{t}^{\mathrm{st}%
})\rightarrow |R).$ On the other hand, notice that the dynamics between
recording events, i.e., Eq.~(\ref{CondicionalPhotonU}), does not affect the
configurational bath state.

From Eqs.~(\ref{sprinkling}) and (\ref{MesureSelectiveConfoGENERAL}), it
follows 
\end{subequations}
\begin{subequations}
\begin{eqnarray}
\digamma _{\mathrm{ph}}[|\rho _{t}^{\mathrm{st}})] &=&\gamma _{R_{t}^{%
\mathrm{st}}}\left\langle +\right\vert \rho _{S}^{\mathrm{st}}(t)\left\vert
+\right\rangle , \\
\digamma _{R}[|\rho _{t}^{\mathrm{st}})] &=&\phi _{RR_{t}^{\mathrm{st}}}.
\end{eqnarray}
Consistently with the classical evolution Eq.~(\ref{ClassicalPconfo}), the
probability by unit of time for observing the configurational transition $%
|R_{t}^{\mathrm{st}})\rightarrow |R)$ [i.e., $\phi _{RR_{t}^{\mathrm{st}}}$]
is independent of the state of the system $\rho _{S}^{\mathrm{st}}(t).$

\begin{figure}[tb]
\includegraphics[bb=26 15 380 560,angle=0,width=7.5
cm]{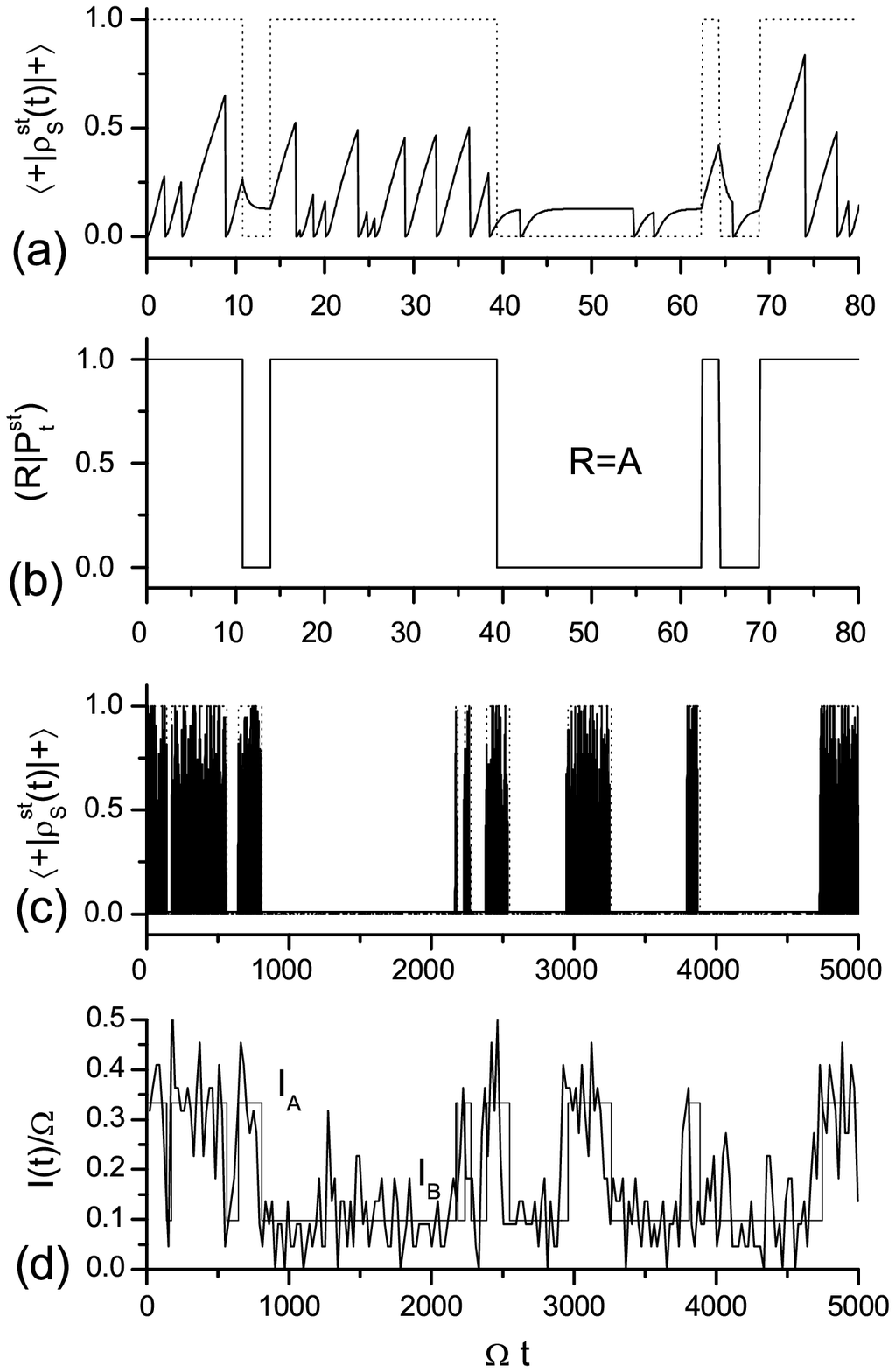}
\caption{Stochastic realizations of a fluorophore system defined by the
evolution Eq.~(\protect\ref{LindbladSelf}) and the measurement operators
Eq.~(\protect\ref{MesureSelectiveConfo}). (a)-(c) Realization of the of the
upper population of the system $\left\langle +\right\vert \protect\rho _{S}^{%
\mathrm{st}}(t)\left\vert +\right\rangle .$ (b) Realization of the of the
configurational population of the bath $(R|P_{t}^{\mathrm{st}}),$ for $R=A.$
(d) Intensity realization. In (a) and (b), the parameters are $\Omega
_{R}=\Omega ,$ $\protect\delta \protect\omega _{R}=0,$ $\protect\gamma %
_{A}/\Omega =1.5,$ $\protect\gamma _{B}/\Omega =3,$ $\protect\phi %
_{AB}/\Omega =0.03,$ $\protect\phi _{BA}/\Omega =0.05,$ and $\protect\omega %
_{L}=\protect\omega _{0}.$ In (c) and (d), the parameters are the same than
in Fig. 1.}
\end{figure}

When a recording event happens, each transformation [Eqs.~(\ref%
{MesureSelectiveConfo})] must to be selected in agreement with the
transition probabilities $\mathrm{t}_{\mu }(t),$ Eq.~(\ref{pu}). They read 
\end{subequations}
\begin{subequations}
\label{trancisionesSelectivasConfo}
\begin{eqnarray}
\mathrm{t}_{\mathrm{ph}}(t) &=&\frac{\gamma _{R_{t}^{\mathrm{st}%
}}\left\langle +\right\vert \rho _{S}^{\mathrm{st}}(t)\left\vert
+\right\rangle }{\gamma _{R_{t}^{\mathrm{st}}}\left\langle +\right\vert \rho
_{S}^{\mathrm{st}}(t)\left\vert +\right\rangle +\tilde{\phi}_{R_{t}^{\mathrm{%
st}}}},  \label{transitionPhoton} \\
\mathrm{t}_{R}(t) &=&\frac{\phi _{RR_{t}^{\mathrm{st}}}}{\gamma _{R_{t}^{%
\mathrm{st}}}\left\langle +\right\vert \rho _{S}^{\mathrm{st}}(t)\left\vert
+\right\rangle +\tilde{\phi}_{R_{t}^{\mathrm{st}}}},  \label{transitionR}
\end{eqnarray}
where $\tilde{\phi}_{R_{t}^{\mathrm{st}}}$\ follows from Eq.~(\ref%
{PhiNormalizadaR}). Notice that when a configurational transition happens, $%
|R_{t}^{\mathrm{st}})\rightarrow |R),$ the different possible final states $%
|R)$ are chosen with probabilities $\mathrm{t}_{R\leftarrow R_{t}^{\mathrm{st%
}}}\equiv \mathrm{t}_{R}(t)/\sum_{R^{\prime }}\mathrm{t}_{R^{\prime
}}(t)=\phi _{RR_{t}^{\mathrm{st}}}/\tilde{\phi}_{R_{t}^{\mathrm{st}}}.$ This
result can straightforwardly be read from the classical master equation~(\ref%
{ClassicalPconfo}).

In Fig.~11 we show the realizations associated to the measurement
transformations Eq.~(\ref{MesureSelectiveConfo}) and the evolution Eq.~(\ref%
{LindbladSelf}). They were build up by using the finite time step algorithm
(Appendix D).

Fig.~11(a) shows a realization of the upper population of the system $%
\left\langle +\right\vert \rho _{S}^{\mathrm{st}}(t)\left\vert
+\right\rangle $ [Eq.~(\ref{SystemConfiSTOCH})]. In contrast with Fig.~1,
here each event may corresponds to a photon detection event, $\left\langle
+\right\vert \rho _{S}^{\mathrm{st}}(t)\left\vert +\right\rangle \rightarrow
0$ [Eq.~(\ref{colapso})], or to a configurational transition (vertical
dotted lines), Eq.~(\ref{transicion}). In these last events the upper
population remains unaffected. In Fig.~11(b), we show the realization of $%
(R|P_{t}^{\mathrm{st}}),$ for $R=A.$ In contrast with Fig.~1, here at all
times we know with total certainty $[(R|P_{t}^{\mathrm{st}})=1$ or $0]$ the
configurational state of the bath.

In Fig.~11(c) and (d) we show the realization of $\left\langle +\right\vert
\rho _{S}^{\mathrm{st}}(t)\left\vert +\right\rangle $ and the scattered
intensity $I(t).$ The parameters are the same than in Fig.~1. In (d), the
telegraphic signal correspond to $\sum_{R}I_{R}(R|P_{t}^{\mathrm{st}}),$
where the intensities $\{I_{R}\}$ are defined by Eq.~(\ref{IntR}). This
function assume the value $I_{R}$ when the bath is in the configurational
state $R.$ The plot shows the direct correlation between the value of the
intensity and the configurational bath state.

\subsubsection{Recording process}

Given that a recording event happens in the $\mu $-detector at time $%
t^{\prime },$ the waiting time distribution for the next event at time $t$
is given by $w_{\mathrm{st}}(t,t^{\prime },\mu ),$ Eq.~(\ref{Wst_mu}). The
event at time $t$ is selected with probabilities $\mathrm{t}_{\mu }(t),$
Eq.~(\ref{trancisionesSelectivasConfo}). Here, $w_{\mathrm{st}}(t,t^{\prime
},\mu )$ can be expressed in a shorter way through its associated survival
probability, i.e., $w_{\mathrm{st}}(t,t^{\prime },\mu
)=-(d/dt)P_{0}[t,t^{\prime };\mathcal{\hat{M}}_{\mu }|\rho _{t^{\prime }}^{%
\mathrm{st}})],$ Eq.~(\ref{WestocasticaPoDef}). From Eqs.~(\ref%
{CondicionalPhotonU}) and (\ref{MesureSelectiveConfo}), we get 
\end{subequations}
\begin{equation}
P_{0}[t,t^{\prime };\mathcal{\hat{M}}_{\mathrm{ph}}|\rho _{t^{\prime }}^{%
\mathrm{st}})]=e^{-\tilde{\phi}_{R_{t^{\prime }}^{\mathrm{st}}}(t-t^{\prime
})}W_{R_{t^{\prime }}^{\mathrm{st}}}[t-t^{\prime };\!\left\vert
-\right\rangle \left\langle -\right\vert ],  \label{PEstoPhoton}
\end{equation}%
and for $R=1,2,\cdots ,R_{\max },$%
\begin{equation}
P_{0}[t,t^{\prime };\mathcal{\hat{M}}_{R}|\rho _{t^{\prime }}^{\mathrm{st}%
})]=e^{-\tilde{\phi}_{R}(t-t^{\prime })}W_{R}[t-t^{\prime };\rho _{S}^{%
\mathrm{st}}(t^{\prime })].  \label{PEstoConfigurationalTransition}
\end{equation}%
Here, $W_{R}[t;\rho ]\equiv \mathrm{Tr}_{S}[\tilde{\rho}_{R}^{\mathrm{u}%
}(t)],$ where $\tilde{\rho}_{R}^{\mathrm{u}}(t)$ is the solution of the
equation $(d/dt)\tilde{\rho}_{R}^{\mathrm{u}}(t)=-(i/\hbar )[H_{R},\tilde{%
\rho}_{R}^{\mathrm{u}}(t)]-\gamma _{R}\{D,\tilde{\rho}_{R}^{\mathrm{u}%
}(t)\}_{+},$ solved with the initial condition $\tilde{\rho}_{R}^{\mathrm{u}%
}(0)=\rho .$ Thus, $W_{R}[t;\rho ]$ is the photon survival probability of a
Markovian system that begins in the state $\rho ,$ and whose characteristic
parameters are $\gamma _{R},$ $\omega _{R},$ and $\Omega _{R}.$ In fact, the
waiting time distribution Eq.~(\ref{WaitingMarkovTiempo}) can also be
written as $w_{R}(t)=-(d/dt)W_{R}[t;\left\vert -\right\rangle \left\langle
-\right\vert ].$

The interpretation of the survival probabilities Eqs.~(\ref{PEstoPhoton})
and (\ref{PEstoConfigurationalTransition}) is very simple. The exponential
factors take into account the\ probability of not having any configurational
transition in the time interval $(t^{\prime },t).$ On the other hand, the
factors defined by $W_{R}[t;\rho ]$ measure the probability of not having
any photon emission in $(t^{\prime },t).$ In Eq.~(\ref%
{PEstoConfigurationalTransition}), $W_{R}[t-t^{\prime };\rho _{S}^{\mathrm{st%
}}(t^{\prime })]$ is the photon survival probability of a Markovian system
(with parameters corresponding to the configurational state $R$) that begins
in the (arbitrary) state $\rho _{S}^{\mathrm{st}}(t^{\prime }).$
Consistently, in Eq.~(\ref{PEstoPhoton}) the factor $W_{R_{t^{\prime }}^{%
\mathrm{st}}}[t-t^{\prime };\left\vert -\right\rangle \left\langle
-\right\vert ]$ corresponds to the photon survival probability after
happening a photon detection event at time $t^{\prime },$ i.e., $\rho _{S}^{%
\mathrm{st}}(t^{\prime })=\left\vert -\right\rangle \left\langle
-\right\vert .$ Hence, here the associated stochastic waiting time
distributions $\{w_{\mathrm{st}}(t,t^{\prime },\mu )\}$\ change when a
configurational transition or when a photon recording event happen. Added to
its dependence on the configurational state $R,$ in contrast with the result
of Section IV, $w_{\mathrm{st}}(t,t^{\prime },\mu )$ also may depends on the
system state $\rho _{S}^{\mathrm{st}}(t^{\prime }),$ i.e., its functional
form depends parametrically on the matrix elements of $\rho _{S}^{\mathrm{st}%
}(t^{\prime }).$

We have checked that the stochastic dynamic of $|\rho _{t^{\prime }}^{%
\mathrm{st}})$ defined by Eqs.~(\ref{PEstoPhoton}) and (\ref%
{PEstoConfigurationalTransition}), like in Fig.~2, also recover the density
matrix evolution defined by Eq.~(\ref{LindbladSelf}). As the dynamic of the
configurational states is classical, the statistical properties of the
photon-emission process remain the same. This fact is clearly seen by
comparing Fig.~1(d) and Fig.~11(d). In both cases the intensity is
characterized by the same telegraphic behavior.

\subsection{Light assisted processes}

Here we analyze the quantum-jump approach associated to Eq.~(\ref%
{LigthMaster}) when both the photon emissions and the configurational
transitions are measured, Eq.~(\ref{single}).

\subsubsection{Measurement operators}

From Eq.~(\ref{LigthMaster}) the measurement transformations read 
\begin{subequations}
\label{MUAssisted}
\begin{eqnarray}
\mathcal{\hat{M}}_{\mathrm{ph}}|\rho ) &=&\frac{\sum_{R}\gamma _{R}|R)\sigma
\rho _{R}\sigma ^{\dagger }}{\sum_{R^{\prime }}\gamma _{R^{\prime }}\mathrm{%
Tr}_{S}[\sigma ^{\dagger }\sigma \rho _{R^{\prime }}]}, \\
\mathcal{\hat{M}}_{R}|\rho ) &=&\frac{|R)\sum_{R^{\prime }}\gamma
_{RR^{\prime }}\sigma \rho _{R^{\prime }}\sigma ^{\dagger }}{\sum_{R^{\prime
\prime }}\gamma _{RR^{\prime \prime }}\mathrm{Tr}_{S}[\sigma ^{\dagger
}\sigma \rho _{R^{\prime \prime }}]},
\end{eqnarray}%
$R\in (1,R_{\max }),$ while the conditional evolution here is also defined
by Eq.~(\ref{Dligth}).

\subsubsection{Stochastic dynamics}

The transformations Eq.~(\ref{MUAssisted}) imply the transformations $[|\rho
_{t}^{\mathrm{st}})=\rho _{S}^{\mathrm{st}}(t)|R_{t}^{\mathrm{st}})]$%
\end{subequations}
\begin{subequations}
\label{MesureSelectiveLight}
\begin{eqnarray}
|\rho _{t}^{\mathrm{st}}) &\rightarrow &\mathcal{\hat{M}}_{\mathrm{ph}}|\rho
_{t}^{\mathrm{st}})=\left\vert -\right\rangle \left\langle -\right\vert
|R_{t}^{\mathrm{st}}), \\
|\rho _{t}^{\mathrm{st}}) &\rightarrow &\mathcal{\hat{M}}_{R}|\rho _{t}^{%
\mathrm{st}})=\left\vert -\right\rangle \left\langle -\right\vert |R).
\end{eqnarray}%
While $\mathcal{\hat{M}}_{\mathrm{ph}}$ collapse the system to its ground
state and leaves invariant the configurational state, the superoperators $%
\mathcal{\hat{M}}_{R}$ produces both the system collapse and the
configurational transition $R_{t}^{\mathrm{st}}\rightarrow R.$ Therefore,
here any recording event (due to $\mathcal{\hat{M}}_{\mathrm{ph}}$ or to $%
\mathcal{\hat{M}}_{R}$) implies a photon detection event.

The transformations defined by Eq.~(\ref{MesureSelectiveLight}) must to be
selected in agreement with the transition probabilities $\mathrm{t}_{\mu
}(t),$ Eq.~(\ref{pu}). From Eqs.~(\ref{sprinkling}) and (\ref{MUAssisted}),
we obtain 
\end{subequations}
\begin{subequations}
\begin{eqnarray}
\digamma _{\mathrm{ph}}[|\rho _{t}^{\mathrm{st}})] &=&\gamma _{R_{t}^{%
\mathrm{st}}}\left\langle +\right\vert \rho _{S}^{\mathrm{st}}(t)\left\vert
+\right\rangle , \\
\digamma _{R}[|\rho _{t}^{\mathrm{st}})] &=&\gamma _{RR_{t}^{\mathrm{st}%
}}\left\langle +\right\vert \rho _{S}^{\mathrm{st}}(t)\left\vert
+\right\rangle .
\end{eqnarray}%
Then, the transition probabilities read 
\end{subequations}
\begin{equation}
\mathrm{t}_{\det }(t)=\frac{\gamma _{R_{t}^{\mathrm{st}}}}{\tilde{\gamma}%
_{R_{t}^{\mathrm{st}}}},\ \ \ \ \ \ \ \ \ \mathrm{t}_{R}(t)=\frac{\gamma
_{RR_{t}^{\mathrm{st}}}}{\tilde{\gamma}_{R_{t}^{\mathrm{st}}}}.
\label{transitionLuz}
\end{equation}%
Notice that these objects are independent of the state $\rho _{S}^{\mathrm{st%
}}(t).$ Furthermore, they are stretched related with the definitions Eq.~(%
\ref{q_erres}).

Fig.~12 shows the realizations associated to the measurement transformations
Eq.~(\ref{MesureSelectiveLight}) and the evolution Eq.~(\ref{LigthMaster}).
The realizations were determined by using the finite time step algorithm
(Appendix D). The parameters are the same than in Fig.~6.

Fig.~12(a) shows a realization of the upper population of the system $%
\left\langle +\right\vert \rho _{S}^{\mathrm{st}}(t)\left\vert
+\right\rangle $ [Eq.~(\ref{SystemConfiSTOCH})]. The vertical dotted lines
correspond to the time where the configurational transitions happen. In
Fig.~12(b), we show the realization of the configurational population $%
(A|P_{t}^{\mathrm{st}}).$ In contrast with Fig.~11(a) and (b), we notice
that here the configurational transitions are always endowed with a photon
emission. Furthermore, at all times the configurational state of the bath is
known with total certainty $[(R|P_{t}^{\mathrm{st}})=1$ or $0]$ [compare
with Fig.~6(b) and (c)].

\begin{figure}[tb]
\includegraphics[bb=26 15 380 560,angle=0,width=7.5
cm]{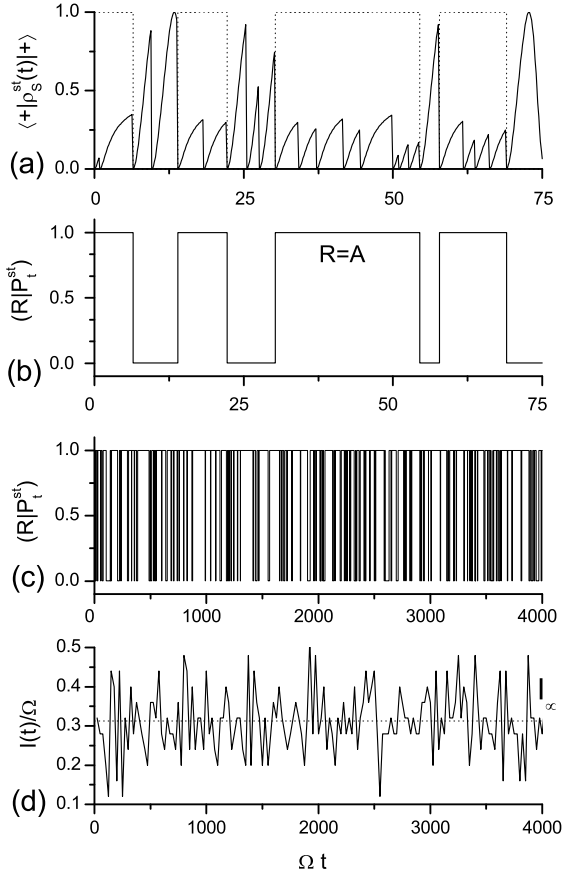}
\caption{Stochastic realizations of a fluorophore system defined by the
evolution Eq.~(\protect\ref{LigthMaster}) and the measurement
transformations Eq.~(\protect\ref{MesureSelectiveLight}). (a) Realization of
the of the upper population of the system $\left\langle +\right\vert \protect%
\rho _{S}^{\mathrm{st}}(t)\left\vert +\right\rangle .$ (b)-(c) Realization
of the of the configurational population of the bath $(R|P_{t}^{\mathrm{st}%
}),$ for $R=A.$ (d) Intensity realization. The parameters are the same than
in Fig.~6.}
\end{figure}

Fig.~12(c) shows $(A|P_{t}^{\mathrm{st}})$ over a larger time scale while in
Fig.~12(d) we show the scattered intensity $I(t).$ As expected, the
intensity realization is similar to that shown in Fig.~6(d).

\subsubsection{Recording process}

The stochastic waiting time distributions Eq.~(\ref{Wst_mu}), from Eqs.~(\ref%
{Dligth}) and (\ref{MesureSelectiveLight}), here read 
\begin{subequations}
\label{WSelectiveLuz}
\begin{eqnarray}
w_{\mathrm{st}}(t,t^{\prime },\mathrm{ph}) &=&\tilde{w}_{R_{t^{\prime }}^{%
\mathrm{st}}}(t-t^{\prime }), \\
w_{\mathrm{st}}(t,t^{\prime },R) &=&\tilde{w}_{R}(t-t^{\prime }),
\end{eqnarray}
\end{subequations}
where $\tilde{w}_{R}(t)$ follows from Eq.~(\ref{WrMarkovTiempo}) after the
replacement $\gamma _{R}\rightarrow \tilde{\gamma}_{R}$ [Eq.~(\ref%
{tildeGamma})], i.e., they are the waiting time distribution of a Markovian
fluorescent system with decay rate $\tilde{\gamma}_{R},$ detuning $\delta
_{R},$ and Rabi frequency $\Omega _{R}. $

The expressions written in Eq.~(\ref{WSelectiveLuz}) only differ in their
sub-index ($R_{t^{\prime }}^{\mathrm{st}}$ or $R$). After a recording event,
the indexes must be chosen with probabilities (\ref{transitionLuz}).
Therefore, $w_{\mathrm{st}}(t,t^{\prime },\mu )$ during successive photon
recording events is randomly selected over the set of functions $\{\tilde{w}%
_{R}(t)\}.$ This result recovers the analysis developed in Ref.~\cite%
{luzAssisted}. For the example shown in Fig.~(12), the two functions $\tilde{%
w}_{A}(t-t^{\prime })$ and $\tilde{w}_{B}(t-t^{\prime })$ can be read from
Fig.~8(a) by taking $p_{A}^{\mathrm{st}}(t^{\prime })=1$ and $p_{A}^{\mathrm{%
st}}(t^{\prime })=0$ respectively. On the other hand, from Eqs.~(\ref%
{transitionLuz}) and (\ref{WSelectiveLuz}), one can deduce that here the
stationary photon waiting time distributions are also defined by Eqs.~(\ref%
{WUnoEstacionariaLigth}) and (\ref{WDOSEstacionariaLigth}).

\section{Stationary n-joint probabilities}

The probabilities Eq.~(\ref{Joint}) define the ensemble statistic of the
measurement process. They depend on the initial condition $\left\vert \rho
_{0}\right) .$ The statistical information that can be obtained from a time
average along a single realization can be obtained from the stationary $n$%
-joint probabilities $P_{n}^{\mathrm{\infty }}\![\tau ,\{\tau
_{i}\}_{1}^{n},\!\{\mu _{i}\}_{1}^{n}].$ They define the events statistics
after happening an infinite number of measurements events and that an
infinite time elapsed since the initial condition, 
\begin{eqnarray*}
P_{n}^{\infty }[\tau ,\{\tau _{i}\}_{1}^{n},\{\mu _{i}\}_{1}^{n}]\! &\equiv
&\!\lim_{N\rightarrow \infty }\lim_{t_{N}\rightarrow \infty
}\int_{0}^{t_{N}}\!\!dt_{N-1}\cdots \!\int_{0}^{t_{2}}\!\!dt_{1} \\
&&\!\sum_{\nu _{N}\cdots \nu _{1}}\!\!\!P_{n+N}[t,\{t_{i}\}_{1}^{n+N},\{\nu
_{i}\}_{1}^{n+N}\!].
\end{eqnarray*}%
The new time variables are defined as $\tau \equiv t-t_{N},$\ $\tau
_{i}\equiv t_{i+N}-t_{N}.$ The measurement apparatus indexes are $\mu
_{i}=\nu _{i+N}.$ By working in a Laplace domain, from Eq.~(\ref{JointCorta}%
) it is possible to obtain 
\begin{eqnarray}
P_{n}^{\mathrm{\infty }}\![\tau ,\{\tau _{i}\}_{1}^{n},\!\{\mu
_{i}\}_{1}^{n}]\!\! &=&\!\!\mathrm{Tr}_{S}[(1|e^{\mathcal{\hat{D}}(\tau
-\tau _{n})}\mathcal{\hat{J}}_{\mu _{n}}\!\!\cdots \!\mathcal{\hat{J}}_{\mu
_{2}}\!e^{\mathcal{\hat{D}}(\tau _{2}-\tau _{1})}  \notag \\
&&\!\times \mathcal{\hat{J}}_{\mu _{1}}e^{\mathcal{\hat{D}}\tau _{1}}%
\mathcal{\hat{M}}|\rho _{\infty })]\digamma \!_{\infty }.
\label{JointEstacionariaCorta}
\end{eqnarray}%
Here the measurement operator $\mathcal{\hat{M}}$ is defined by Eq.~(\ref%
{EME}), and the constant $\digamma \!_{\infty }$ reads%
\begin{equation}
\digamma \!_{\infty }\equiv \sum_{\mu }\digamma _{\mu }[|\rho _{\infty })]=%
\mathrm{Tr}_{S}[(1|\mathcal{\hat{J}}|\rho _{\infty })],
\end{equation}%
where $\digamma _{\mu }$ follows from Eq. (\ref{sprinkling}) and we used
Eq.~(\ref{EME}), $\mathcal{\hat{J}}\equiv \sum_{\mu }\mathcal{\hat{J}}_{\mu
}.$ The stationary state $|\rho _{\infty })$\ is defined by Eq.~(\ref%
{RhoInfinity}).

From Eq.~(\ref{JointEstacionariaCorta}), by performing the inverse
calculations steps than in the derivation of Eq.~(\ref{JointCorta}), it
follows%
\begin{eqnarray}
P_{n}^{\infty }[\tau ,\{\tau _{i}\}_{1}^{n},\{\mu _{i}\}_{1}^{n}]
&=&P_{0}[\tau ,\tau _{n};\mathcal{\hat{M}}_{\mu _{n}}\!\left\vert \rho
_{\tau _{n}}\right) ]  \label{JointStationary} \\
&&\prod_{j=2}^{n}w_{\mu _{j}}[\tau _{j},\tau _{j-1};\mathcal{\hat{M}}_{\mu
_{j-1}}\!|\rho _{\tau _{j-1}})]  \notag \\
&&\times w_{\mu _{1}}[\tau _{1},0;\mathcal{\hat{M}}\!|\rho _{\infty
})]\digamma \!_{\infty }.  \notag
\end{eqnarray}%
The auxiliary states read $|\rho _{\tau _{i+1}})=\mathcal{\hat{T}}(\tau
_{i+1},\tau _{i})\mathcal{\hat{M}}_{\mu _{i}}|\rho _{\tau _{i}}),$ where $%
|\rho _{\tau _{1}})=\mathcal{\hat{T}}(\tau _{1},0)\mathcal{\hat{M}}\!|\rho
_{\infty }).$

The interpretation (and structure) of Eq.~(\ref{JointStationary}) is similar
to that of Eq.~(\ref{Joint}). Nevertheless, here the factor $\digamma
\!_{\infty }$ takes into account the probability by unit of time of having
an arbitrary detection event in the long time regime. The associated
measurement operator is $\mathcal{\hat{M}},$ Eq.~(\ref{EME}). Furthermore,
in contrast to Eq.~(\ref{Joint}), the first contribution (waiting time
distribution) in Eq.~(\ref{JointStationary}) is defined with the state $%
\mathcal{\hat{M}}|\rho _{\infty }),$ i.e., the state after and arbitrary
detection happening in the stationary regime. From Eqs.~(\ref%
{JointStationary}) and (\ref{JointEstacionariaCorta}), the expressions Eqs.~(%
\ref{waitST1}) and (\ref{waitST2}) follows straightforwardly after replacing 
$\tau _{1}\rightarrow \tau _{1},$ and $\tau _{2}\rightarrow \tau _{1}+\tau
_{2}.$ In fact, the variables $\{\tau _{i}\}_{1}^{n}$\ of the stationary
waiting time distributions $w_{\infty }^{(n)}\![\{\tau
_{i}\}_{1}^{n},\!\{\mu _{i}\}_{1}^{n}]$ denotes the time interval between
consecutive recording events.

\section{Averaging over realizations}

Here, we demonstrate that the deterministic evolution Eq.~(\ref{vectorial})
is recovered after averaging Eq.~(\ref{stochastic}) over realizations of the
Poisson processes $N_{t}^{\mu }.$

First, by using that $(dN_{t}^{\mu })^{k}=dN_{t}^{\mu }$ and the property $%
dN_{t}^{\mu }dN_{t}^{\mu ^{\prime }}=\delta _{\mu \mu ^{\prime }}dN_{t}^{\mu
},$ it is possible to get the relation \cite{breuerbook} 
\begin{equation}
\overline{\Xi (\{N_{t}^{\mu }\})dN_{t}^{\mu }}=\overline{\Xi (\{N_{t}^{\mu
}\})\mathrm{Tr}_{S}[(1|\mathcal{\hat{J}}_{\mu }|\rho _{t}^{\mathrm{st}})]\ }%
dt,  \label{AveragePoisson}
\end{equation}%
where $\Xi (\{N_{t}^{\mu }\})$ is an arbitrary function of the Poisson
processes $\{N_{t}^{\mu }\}.$ This equality can be immediately deduced by
introducing a series expansion of $\Xi .$ Now, we split the average of Eq.~(%
\ref{stochastic}) as%
\begin{equation}
\frac{d}{dt}|\rho _{t})=\left. \frac{d}{dt}|\rho _{t})\right\vert _{\mathcal{%
\hat{D}}}+\left. \frac{d}{dt}|\rho _{t})\right\vert _{\mathcal{\hat{M}}},
\end{equation}%
where the first contribution is associated to the conditional deterministic
dynamics and the second one with the disruptive measurement changes. Then,
trivially it follows%
\begin{equation}
\left. \frac{d}{dt}|\rho _{t})\right\vert _{\mathcal{\hat{D}}}=\mathcal{\hat{%
D}}|\rho _{t})-\overline{\mathrm{Tr}_{S}[(1|\mathcal{\hat{D}}|\rho _{t}^{%
\mathrm{st}})]|\rho _{t}^{\mathrm{st}})}.  \label{DerivadaConditional}
\end{equation}%
On the other hand, by using the definition Eq.~(\ref{measurement}) and the
relation Eq.~(\ref{AveragePoisson}), we get%
\begin{equation}
\left. \frac{d}{dt}|\rho _{t})\right\vert _{\mathcal{\hat{M}}}=\mathcal{\hat{%
J}}_{\mu }|\rho _{t})-\sum_{\mu }\overline{\mathrm{Tr}_{S}[(1|\mathcal{\hat{J%
}}_{\mu }|\rho _{t}^{\mathrm{st}})]|\rho _{t}^{\mathrm{st}})\ }.
\label{DerivadaJump}
\end{equation}%
After introducing the relation $\mathrm{Tr}_{S}[(1|\mathcal{\hat{D}}|\bullet
)]=-\sum_{\mu }\mathrm{Tr}_{S}[(1|\mathcal{\hat{J}}_{\mu }|\bullet )]$ in
Eqs.~(\ref{DerivadaConditional}) and (\ref{DerivadaJump}), the evolution
Eq.~(\ref{vectorial}) follows straightforwardly.

\section{Algorithms associated to the stochastic evolution}

Two different algorithms allow to build up the realizations associated to
the stochastic evolution Eq.~(\ref{stochastic}).

\subsection{Infinitesimal time step algorithm}

In the first algorithm, the stochastic state $|\rho _{t+\Delta t}^{\mathrm{st%
}})$ is obtained from $|\rho _{t}^{\mathrm{st}}),$ where $\Delta t$ is the
time discretization step. By defining the quantity%
\begin{equation}
\digamma (t)\equiv \sum_{\mu }\digamma _{\mu }[|\rho _{t}^{\mathrm{st}%
})]=\sum_{\mu }\mathrm{Tr}_{S}[(1|\mathcal{\hat{J}}^{\mu }|\rho _{t}^{%
\mathrm{st}})],
\end{equation}%
the probability $\Delta P$ of having a measurement event is defined by $%
\Delta P=\Delta t\ \digamma (t).$ Then, a random number $r$ in $(0,1)$ is
generated and compared with $\Delta P.$ If $r>\Delta P,$ no recording event
happens, so the vectorial state evolves\ deterministically as [Eq.~(\ref%
{ConditionalNormalizada})]%
\begin{equation}
|\rho _{t+\Delta t}^{\mathrm{st}})=\mathcal{\hat{T}}(t+dt,t)|\rho _{t}^{%
\mathrm{st}})\simeq \frac{(1+\mathcal{\hat{D}}\Delta t)|\rho _{t}^{\mathrm{st%
}})}{1+\mathrm{Tr}_{S}[(1|\mathcal{\hat{D}}\Delta t|\rho _{t}^{\mathrm{st}})]%
}.
\end{equation}%
If $r<\Delta P,$ there is measurement event. Then, the system state at $%
t+\Delta t$ is defined by [Eq.~(\ref{measurement})]%
\begin{equation}
|\rho _{t+\Delta t}^{st})=\mathcal{\hat{M}}_{\mu }|\rho _{t}^{st})=\frac{%
\mathcal{\hat{J}}_{\mu }|\rho _{t}^{\mathrm{st}})}{\mathrm{Tr}_{S}[(1|%
\mathcal{\hat{J}}_{\mu }|\rho _{t}^{\mathrm{st}})]}.
\end{equation}%
Here, the index $\mu $\ is chosen with probability $\mathrm{t}_{\mu }(t),$
Eq.~(\ref{pu}). Due to the relation $\digamma _{\mu }[|\rho _{t})]=\overline{%
\mathrm{t}_{\mu }(t)\digamma (t)},$ the generated realizations satisfy Eq.~(%
\ref{stochastic}).

\subsection{Finite time step algorithm}

An alternative and more efficient algorithm can be defined by using the
survival probability Eq.~(\ref{Survival}) [see also Eq.~(\ref%
{WestocasticaPoDef})]. Given that the state of the system after a
measurement at time $t_{i}$ is given by $\mathcal{\hat{M}}_{\mu _{i}}|\rho
_{t_{i}}^{\mathrm{st}}),$\ the time $t_{i+1}$ of the next event is obtained
from the equation%
\begin{equation}
P_{0}[t_{i+1},t_{i};|\rho _{t_{i}}^{\mathrm{st}})]=\mathrm{Tr}_{S}[(1|e^{%
\mathcal{\hat{D}}(t_{i+1}-t_{i})}\mathcal{\hat{M}}_{\mu _{i}}|\rho _{t_{i}}^{%
\mathrm{st}})]=r,  \label{SurvivalRandom}
\end{equation}%
where as before $r$ is a random number in the interval $(0,1).$ For $t\in
(t_{i+1},t_{i}),$ the stochastic state evolves deterministically as [Eq.~(%
\ref{ConditionalNormalizada})]%
\begin{equation}
|\rho _{t}^{\mathrm{st}})=\mathcal{\hat{T}}(t,t_{i})\mathcal{\hat{M}}_{\mu
_{i}}|\rho _{t_{i}}^{\mathrm{st}})=\frac{e^{\mathcal{\hat{D}}(t-t_{i})}%
\mathcal{\hat{M}}_{\mu _{i}}|\rho _{t_{i}}^{\mathrm{st}})}{\mathrm{Tr}%
_{S}[(1|e^{\mathcal{\hat{D}}(t-t_{i})}\mathcal{\hat{M}}_{\mu _{i}}\left\vert
\rho _{t_{i}}^{\mathrm{st}}\right) ]}.  \label{Normalizada}
\end{equation}%
At time $t=t_{i+1},$ an index $\mu _{i+1}$\ is chosen with probability $\{%
\mathrm{t}_{\mu }(t_{i+1})\},$ Eq.~(\ref{pu}), and then the sudden
transformation%
\begin{equation}
|\rho _{t_{i+1}}^{\mathrm{st}})\rightarrow \mathcal{\hat{M}}_{\mu
_{i+1}}|\rho _{t_{i+1}}^{\mathrm{st}})=\frac{\mathcal{\hat{J}}_{\mu
_{i+1}}|\rho _{t_{i+1}}^{\mathrm{st}})}{\mathrm{Tr}_{S}[(1|\mathcal{\hat{J}}%
_{\mu _{i+1}}|\rho _{t_{i+1}}^{\mathrm{st}})]},
\end{equation}%
is applied. The first event follows from Eq.~(\ref{SurvivalRandom}) with $%
\mathcal{\hat{M}}_{\mu _{i}}|\rho _{t_{i}}^{\mathrm{st}})\rightarrow |\rho
_{0}^{\mathrm{st}}).$ The realizations generated with this algorithm are
also consistent with the evolution Eq.~(\ref{stochastic}).



\begin{thebibliography}{99}
\bibitem{VectorNotation} In the context of Ref.~\cite{OpenSMS},\ each vector 
$|R)$ can be associated to a projector $\left\vert R\right\rangle
\left\langle R\right\vert ,$ where the states $\{\left\vert R\right\rangle
\}_{R=1}^{R_{\max }}$ define the effective configurational Hilbert space.
\end{thebibliography}
\end{document}